\documentclass{article}
\usepackage[utf8]{inputenc}
\usepackage[round,numbers]{natbib}
\usepackage{hyperref}
\usepackage{graphicx}
\usepackage[singlespacing]{setspace}
\usepackage{booktabs}
\usepackage{caption}
\usepackage{subcaption}
\usepackage{float}
\usepackage{multirow}
\usepackage{multicol}
\usepackage{adjustbox}
\usepackage[letterpaper,margin=1.1in]{geometry}
\title{Market Potential for CO$_2$ Removal and Sequestration from Renewable Natural Gas Production in California}
\author{\protect\textsc{Jun Wong} \\ \normalsize UC Berkeley, Environmental Science, Policy, and Management \& NYU Stern
\and
\protect\textsc{Jonathan Santoso} \\ \normalsize UC Berkeley, Chemical and Biomolecular Engineering
\and
\protect\textsc{Marjorie Went} \\ \normalsize UC Berkeley, Chemical and Biomolecular Engineering
\and
\protect\textsc{Daniel Sanchez\thanks{Correspoding author: sanchezd@berkeley.edu.}} \\ \normalsize UC Berkeley, Environmental Science, Policy, and Management\\ \\ }
\date{May 2021}

\begin{document}
\singlespacing
\maketitle

\begin{abstract}
Bioenergy with Carbon Capture and Sequestration (BECCS) is critical for stringent climate change mitigation, but is commercially and technologically immature and resource-intensive. In California, state and federal fuel and climate policies can drive first-markets for BECCS. We develop a spatially explicit optimization model to assess niche markets for renewable natural gas (RNG) production with carbon capture and sequestration (CCS) from waste biomass in California. Existing biomass residues produce biogas and RNG and enable low-cost CCS through the upgrading process and CO$_2$ truck transport. Under current state and federal policy incentives, we could capture and sequester 2.9 million MT CO$_2$/year (0.7\% of California's 2018 CO$_2$ emissions) and produce 93 PJ RNG/year (4\% of California's 2018 natural gas demand) with a profit maximizing objective. Existing federal and state policies produce profits of \$11/GJ. Distributed RNG production with CCS potentially catalyzes markets and technologies for CO$_2$ capture, transport, and storage in California. 
\end{abstract}

\vspace{0.5cm}

\doublespacing
\newpage

\section*{Introduction}

Deep decarbonization of the energy system critically relies on Bioenergy with Carbon Capture and Sequestration (BECCS) to produce low-carbon and carbon-negative products such as heat, electricity, and fuels \cite{aines_building_2019}. Current BECCS deployment has been lagging the scales envisaged in century-scale climate change mitigation scenarios. Instead it is limited to a handful of demonstration projects in corn ethanol and waste-to-energy production (\citenum{sanchez_near-term_2018}, \citenum{anderson_trouble_2016}, \citenum{kearns_waste--energy_2019}). Based on this disparity, numerous scholars have argued for a “bottom up” approach to BECCS commercialization based on niche markets, scale-up, regionally appropriate feedstocks, and local policies (\citenum{mollersten_potential_2003}, \citenum{bellamy_govern_2019}, \citenum{nemet_negative_2018}). In these contexts, small-scale (megatonne scale) commercially viable opportunities for BECCS are necessary to promote more widespread adoption of carbon removal. Technology and policy analyses that identify opportunities to leverage existing infrastructure, deploy existing technologies, and capitalize on current policies can enhance mitigation efforts by developing experience in carbon capture, transport, sequestration, and removal \cite{sanchez_near-term_2018}. Moreover, establishing first-markets for BECCS-use can further incentivize and accelerate innovation in this field. We explore such an opportunity in California (CA), producing renewable natural gas with carbon capture and sequestration (RNG-CCS).

\begin{figure}[H]
    \centering
    \caption{Available biomass residues, existing digesters, and potential CO2 sequestration sites in California}
     \label{fig:resources}
    \includegraphics{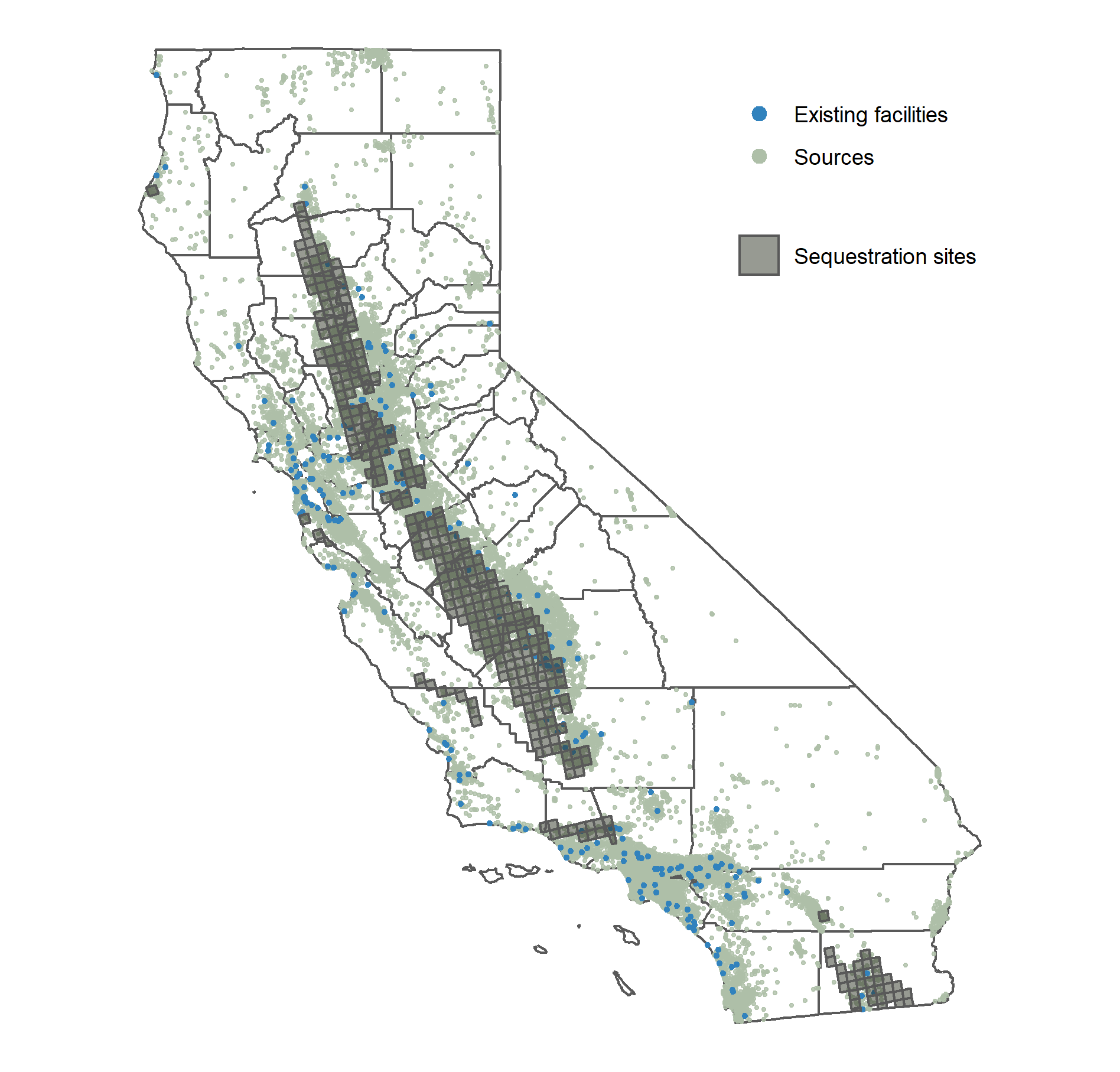}
\end{figure}

California is well-positioned for BECCS implementation. The state’s economy yields large volumes of waste biomass from forestry, urban and agricultural activities. In total, California produced 54 million bone dry tons (BDT) of biomass residues in 2018 \cite{breunig_temporal_2018}.  This is projected to grow to 71 million bdt per year by 2050. At the same time, its geography is well-suited for CO$_2$ storage \cite{baker_getting_2020}. Over 24,000 km$^2$ of non-urban area in the Central Valley, Bay Area, North Coast, and greater Los Angeles area (10\% of California's total land area) is suitable for CO$_2$ sequestration (Figure \ref{fig:resources}). In total, the state has over 200 GtCO$_2$ of total sequestration capacity. We estimate that in-state capacity-weighted sequestration costs are around \$1.35/tCO$_2$ \cite{sanchez_near-term_2018}\footnote{This estimate includes injection and monitoring costs. Please refer to \citet{sanchez_near-term_2018} and the supplmentary information.}. 

Current state and federal fuel and climate policies can drive first-markets for BECCS in California. In particular, the California Low Carbon Fuel Standard (LCFS), federal 45Q carbon sequestration tax credit, and federal Renewable Fuel Standard (RFS) can all be used to incentivize BECCS. Established in 2011, California’s LCFS provides financial incentives for low-carbon and carbon-negative transportation fuels through tradeable CO$_2$ abatement credits \cite{witcover_status_2018}. In 2019, California’s Air Resources Board amended the LCFS to allow for CCS projects to generate additional LCFS credits \cite{california_air_resources_board_low_2019}. Abatement credits are performance-based, proportionate to the carbon intensity reduction relative to a benchmark fossil fuel (E100). Historical credit prices ranged from \$160-192 tCO$_2$ abated between 2018-2020. In 2018, the federal government enhanced its existing tax credit for geologic CO$_2$ sequestration, providing 45Q tax credits as high as \$50/tCO$_2$ sequestered for dedicated storage for facilities capturing over 100,000 tCO$_2$ per year. In addition, the federal RFS subsidizes the production of renewable fuels through the distribution of Renewable Identification Number (RIN) credits. Various biofuels are eligible to receive different classes of RIN credits on a volume-basis (one credit is a Gallon Gasoline Equivalent (GGE)): lignocellulosic biofuels receive D3 credits, which often receive higher prices, while “advanced” (non-corn starch) biofuels receive D5 credits. D3 credits have traded at 0.59-2.74 \$/GGE in 2018-2020, while D5 credits have traded at 0.38-0.85 \$/GGE.

In recent years, California’s waste management policies have been in favor of biomass utilization. SB 1383, signed into law in 2016, is a broad methane reduction strategy in California that targets a 40\% decrease in methane emissions by 2030 and 75\% decrease in organic waste sent to landfills by 2025. The bill specifically targets dairy emissions: it requires a 40\% decrease of dairy production-related methane emissions by 2030. Beyond dairy, the legislation targets a 50\% reduction in the level of the statewide disposal of organic waste from the 2014 level by 2020 and a 75\% reduction by 2025 \cite{lara_bill_nodate}. SB1383 necessitates innovative methods to process waste biomass. Anaerobic digestion, electricity generation, biochar production for land application, and composting are all methods that can utilize the available biomass residues and potentially meet the SB1383 goals.  Of these applications, RNG-CCS through anaerobic digestion is a particularly attractive candidate to meet both the organic waste diversion and methane emissions goals while simultaneously producing carbon-negative fuels. 

Given the policy and resource context, California has seen large growth in biogas and renewable natural gas production from biomass via anaerobic digestion and landfill gas collection. Anaerobic digestion, a sequence of processes by which microorganisms break down biodegradable material in the absence of oxygen, has been used as a technology for waste management and renewable fuel production. We estimate that 46 anaerobic digester (AD) sites process food waste, municipal solid waste, and dairy waste in California as of 2019 (\citenum{us_epa_agstar_2014}, \citenum{argonne_national_laboratory_renewable_nodate}, \citenum{calrecycle_california_2019}). Furthermore, 154 existing wastewater treatment plants process wastewater and 52 landfills are equipped with existing landfill gas (LFG) collection systems \cite{us_epa_landfill_2016}. We estimate roughly 3 million mmbtu of biogas is produced per year from anaerobic digesters and 10 million mmbtu of landfill gas is diverted to various cogeneration projects in California. All but five AD projects in California are generating electricty and/or heat.  More recently, producers are starting to upgrade biogas into renewable natural gas (RNG) for injection into California’s natural gas distribution system. RNG has entered markets as a low-carbon transportation fuel, in part due to subsidies from the RFS and LCFS. For instance, the Calgren facility in Pixley and the CR\&R facility in Perris generate RNG from manure and food waste, respectively. 

Despite the recent development of RNG facilities in California, the several existing facilities in California do not incorporate RNG production with CCS despite the fertile policy environment and geologic endowment in California.  Producing RNG from biogas presents an opportunity for low-cost CO$_2$ capture. Biogas contains a mixture of CH$_4$ and CO$_2$, along with other trace contaminants. The upgrading process separates CH$_4$ from other components of biogas, producing not only a pure stream of CH$_4$, but also a relatively pure stream of CO$_2$. In particular, Pressure Swing Adsorption (PSA) separates the CH$_4$ and CO$_2$ in biogas and produces a stream of 75-98\% pure CO$_2$ (\citenum{ong_comparative_2014}, \citenum{sun_selection_2015}). To our knowledge, RNG-CCS is relatively under-studied in the literature despite the attractive low-cost opportunity for CO$_2$ capture. Most recently, \citet{esposito_simultaneous_2019} reports that the production of food-grade CO$_2$ from biogas upgrading (99.9\%) can be profitable in Italy.

\begin{figure}[H]
    \caption{System and Model Flow Chart}
    \label{fig:flowchart}
    \begin{subfigure}{\textwidth}
        \caption{System Flow Chart}
        \centering
        \includegraphics[width=0.8\linewidth]{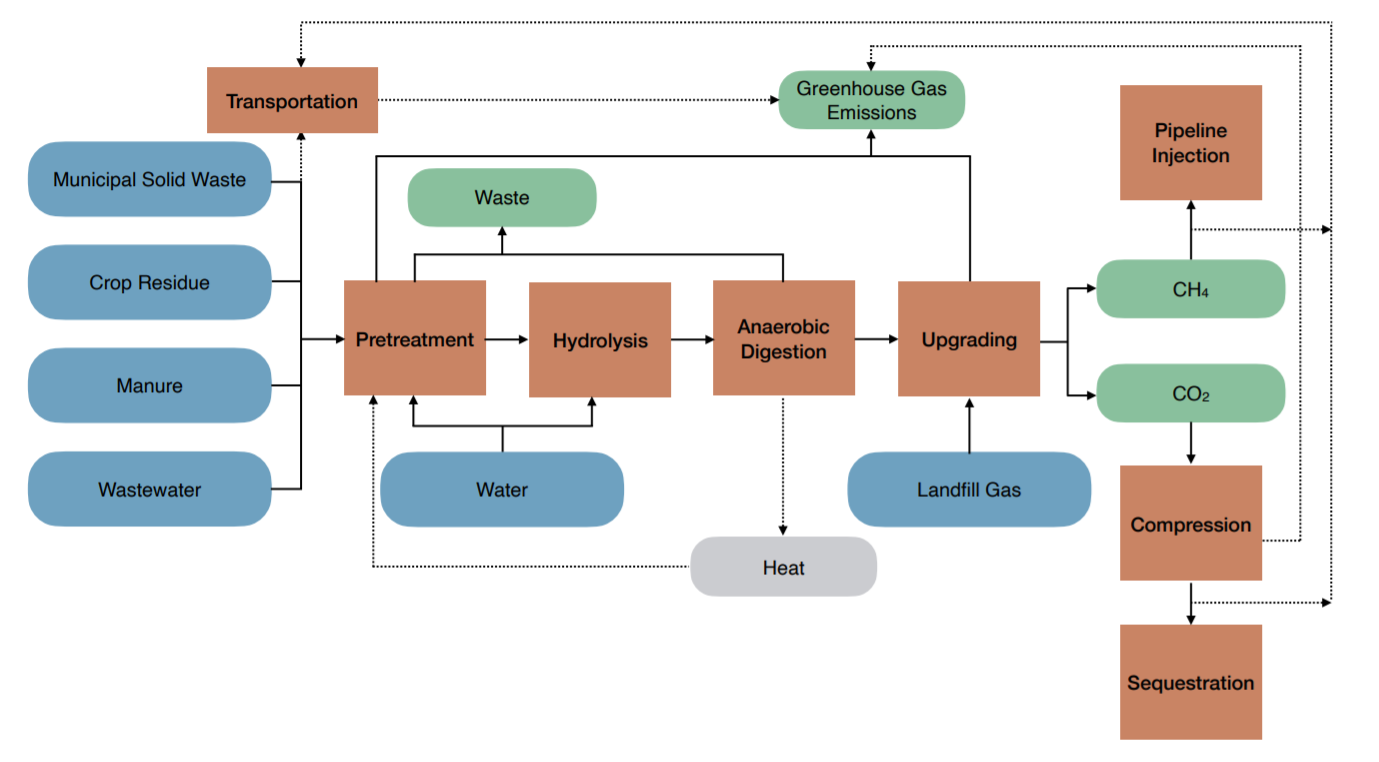}
    \end{subfigure}
    \begin{subfigure}{\textwidth}
        \caption{Model Flow Chart}
        \centering
        \includegraphics[width=0.8\linewidth]{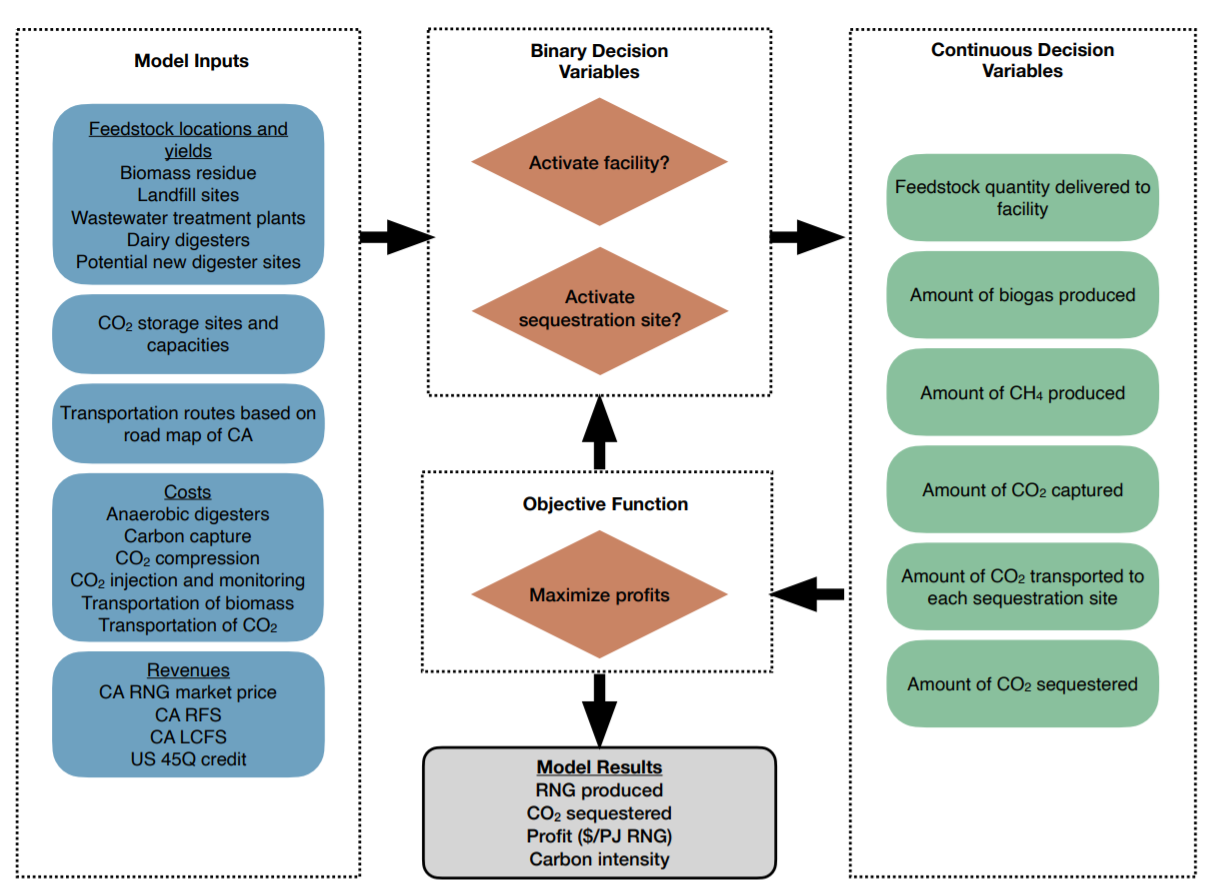}
    \end{subfigure}
\end{figure}

Here, we develop a spatially explicit optimization model to assess the near term opportunities for RNG production from waste biomass with CCS in California under existing policy incentives (Figure \ref{fig:flowchart}). We incorporate high-resolution data on existing biomass residues, anaerobic digesters, wastewater treatment plants, landfills, truck networks, and geologic sequestration sites in California. Profitable RNG production with CCS has the potential to catalyze markets and technologies for CCS and BECCS in California.

\section*{Methods}

\subsection*{Data Development}
We compile a database of available biomass residues (feedstocks) in California from various sources. Data on municipal solid waste (MSW), crop residues, and manure are from \citet{breunig_temporal_2018}. We supplement these with data on wastewater treatment plants and daily average flows from the California Association of Sanitation Agencies (CASA). Lastly, we collect landfill locations and volumes of landfill gas from EPA Landfill Methane Outrech Program (LMOP). We estimate the current deployment of anaerobic digesters in California using the EPA AgStar database on dairy digesters \cite{us_epa_agstar_2014}, Argonne National Lab RNG database \cite{argonne_national_laboratory_renewable_nodate}, EPA LMOP database \cite{us_epa_landfill_2016}, and CalRecycle \cite{calrecycle_california_2019}. We consider locations for new digesters in cities with population greater than 10,000 and less than 5 km away from an existing natural gas pipelines. Saline aquifer storage capacities and locations are derived from v1502 of the National Carbon Sequestration Database (SI Text) \cite{noauthor_natcarbatlas_nodate}. We use the open source routing machine \cite{noauthor_project_nodate} to calculate the shortest road distance and time between point sources and facilities, and facilities to sequestration sites. To avoid the computational limitations, we limit feedstock and CO$_2$ transport to within 50 miles.  

We derive the cost of anaerobic digesters, biogas upgrading, and RNG injection from \citet{parker_renewable_2017} (SI text). We use the biomass transport cost provided in \citet{tittmann_spatially_2010}. We use the cost of CO$_2$ capture and transport provided in \citet{psarras_carbon_2017}, and CO$_2$ compression and pumping cost from \citet{mccollum_techno-economic_2006}. To calculate the costs of CO$_2$ sequestration in saline aquifers, we supplement the methods of \citet{sanchez_near-term_2018} with the C2SAFE model for California-specific policy related costs, such as seismic monitoring \cite{trautz_california_2018}. We convert all costs to 2019 US dollars using the Information Handling Services (IHS) Upstream Capital Cost Index. We assume a 15-year project timeline and 10\% internal rate of return. We assume that all the RNG produced is used to supply compressed natural gas (CNG) in California as transportation fuel. Thus, the RNG is eligible to receive Renewable Identification Numbers (RINs) under the Renewable Fuels Standard. The RNG produced is also eligible to receive the Low Carbon Fuel Standard (LCFS) credits and 45Q carbon sequestration credits (given the 100,000 tCO$_2$ captured per year threshold is met), and the market price for CNG.

We mainly use biogas yields from \citet{li_comparison_2013}, but we also compile a range of biogas yields for various feedstocks from a broader literature investigation (SI Text). We assume that the resulting biogas is comprised of 60\% CH$_4$ and 40\% CO$_2$. To determine the amount of LCFS credits, we take the carbon intensity of the RNG produced from the California Air Resources Board \cite{california_air_resources_board_low_2019}. Additional sequestered CO$_2$ also earn LCFS credits. We assume that the system uses California grid electricity for capture and compression, with carbon intensity equal to the average California mix in 2018 \cite{us_epa_emissions_2015}. We also assume for CO$_2$ transport an emissions factor of 161.8 gCO$_2$/ton-mile \cite{environmental_defense_fund_green_nodate}. Note that the feedstock transport emissions factor is already accounted for in the GHG intensity provided by the Air Resources Board.  

We build a spatially explicit optimization model informed by data on biomass residues, existing anaerobic digesters, wastewater treatment plants, landfills, road networks, and geologic sequestration sites in California (Figure \ref{fig:flowchart}). In this paper, we offer two novel methodological contributions to the literature on spatial optimization of biomass and CCS. We model for codigestion and consider trucking networks for CO$_2$ transportation. We contend that the codigestion of feedstocks offers greater economies of scale and flexibility to the system. Additionally, the relatively small amounts of CO$_2$ produced in this context do not justify pipeline construction for the sole purpose of transporting several million tons of CO$_2$. In particular, California’s siting and permitting processes are not conducive for near-term development of CO$_2$ pipelines. 

\subsection*{Problem Statement and Scenarios}
We minimize the net total cost of the production of RNG and sequestration of CO$_2$ using mixed integer programming to identify cost-effective sequestration opportunities (Figure \ref{fig:flowchart}). We define two binary decision variables that indicate the activation of facilities and sequestration sites, respectively. The six continuous decision variables are: the quantity of feedstock delivered to facility, amount of biogas produced, amount of CO$_2$ captured from the biogas upgrading process, amount of CH$_4$ produced from the biogas upgrading process, amount of CO$_2$ transported to each sequestration site and the amount of CO$_2$ sequestered at each sequestration site. Our model is implemented in AMPL and solved using a branch-and-bound method. The model’s complexity yields optimality gaps of 10-20\%--implying that solutions are feasible but not necessarily optimal. 

Within each scenario, we assume a range of market price of RNG, and  credit prices of LCFS, RIN, and 45Q. We detail the system boundaries in the SI. In practice, price volatility, policy uncertainty, lifecycle emissions accounting, and tax equity availability will all affect the cost-effectiveness of CCS projects that is dependent on tax credits and tradeable permits. We do not explicitly model for other potential revenue streams, such as CO$_2$ utilization options (EOR or beverage carbonation) or alternative feedstock uses. 

\subsection*{Data Access}

Existing pipeline rights of way are available \href{https://www.npms. phmsa.dot.gov/}{here}. National Carbon Sequestration database on saline aquifers are available \href{https://www.netl.doe.gov/research/coal/carbon-storage/ natcarb-atlas}{here}. The EPA Landfill and Outreach Program database is available \href{https://www.epa.gov/lmop}{here}. Existing agricultural digester database in the U.S. can be found at the EPA AgStar database, available here \href{https://www.epa.gov/agstar}{here}. Our model and a reproduction kit is available on \href{https://github.com/carbon-removal-laboratory/Biogas-CCS}{GitHub}.

\section*{Results and Discussion}

\begin{figure}[H]
\caption{Baseline Results}
\label{fig:baseline}
\begin{subfigure}{\textwidth}
    \caption{California}
        \centering
    \includegraphics[width=0.8\linewidth]{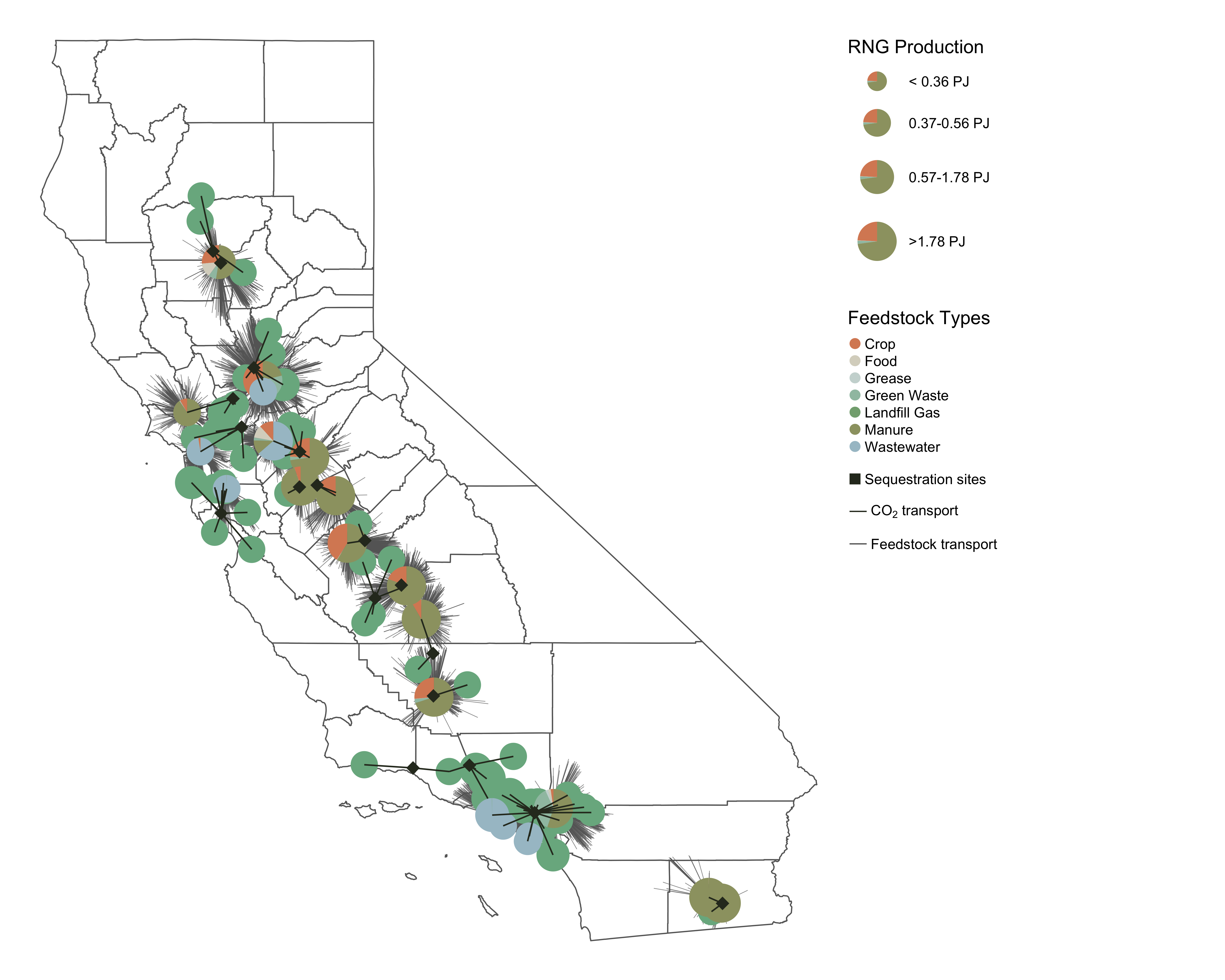}

\end{subfigure}

\begin{subfigure}{\textwidth}
    \caption{Sacramento Area}
        \centering
    \includegraphics[width=0.6\linewidth]{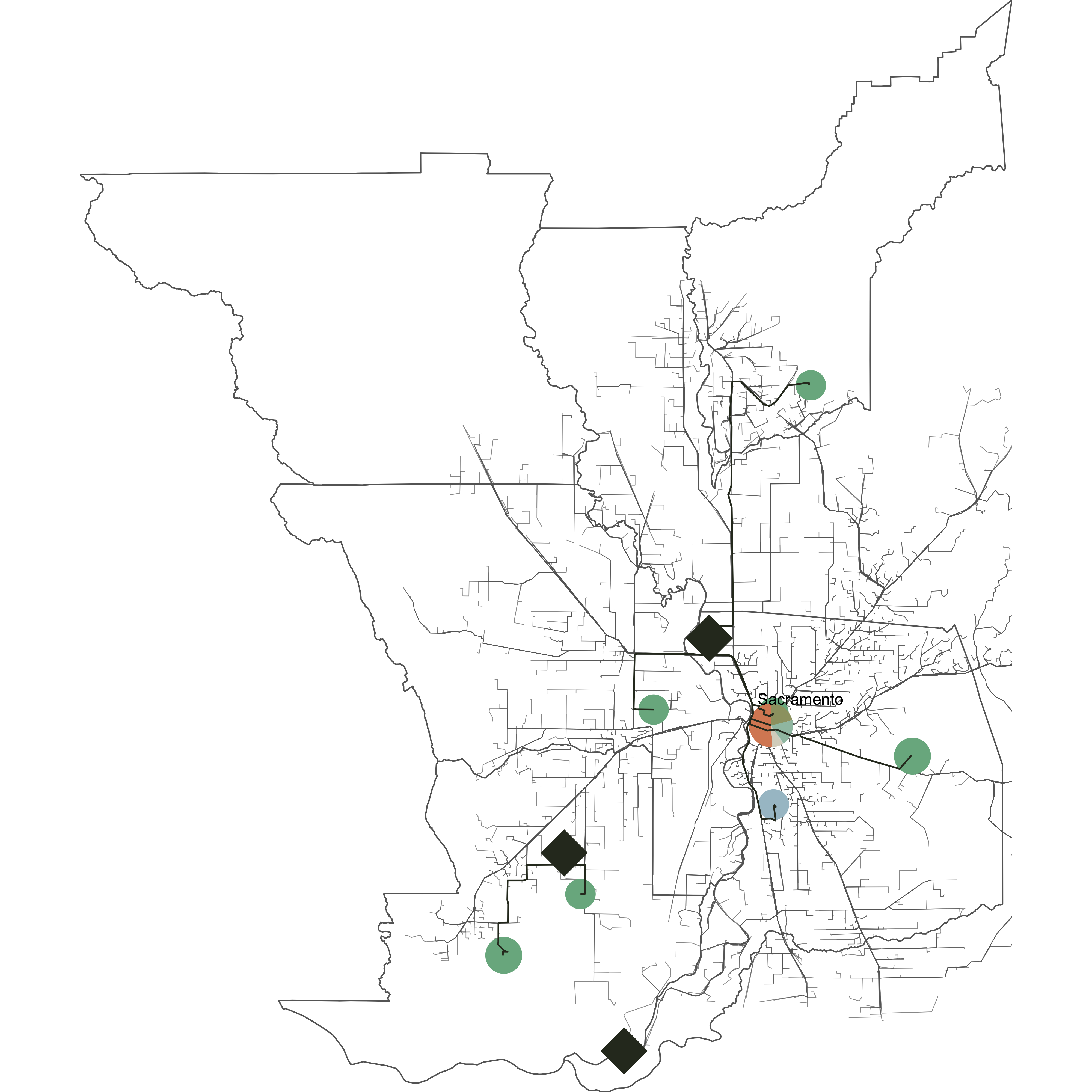}

\end{subfigure}
\end{figure}

To characterize the optimal deployment of RNG production with CCS in California (RNG-CCS) under current policy conditions, we present key model outputs—RNG produced, CO$_2$ sequestered, costs, revenues, and profits—under a range of policy scenarios (Figure \ref{fig:baseline}). Policy scenarios are constructed around recent prices available under Section 45Q, California’s LCFS, and the federal RFS (Table 1). Further information about our methods and assumptions is available in the Methods and SI Text.

California’s economy yields large volumes of waste biomass--we use as basis the estimated 54 billion bdt of biomass residues available from crop residues, municipal waste, and manure per year in 2020 \cite{breunig_temporal_2018}. Additionally, exisiting landfill gas facilities and wastewater treatment facilities collect 5.3 million m$^3$ landfill gas and 2.82 million gallons of wastewater per day, respectively. Taken together, these sources can produce up to 146 PJ RNG per year, or 6.5\% of the natural gas demand in California in 2018, and 5 million tons CO$_2$ per year for potential sequestration \cite{eia2020rng}.

In the baseline scenario, 79 facilities profitably produce 93 PJ of RNG per year (a near ten-fold increase over present levels) and sequester 2.9 million tons of CO$_2$ per year in 18 sequestration sites. The volume of CO$_2$ sequestered at each site ranges up to 515,000 tons CO$_2$ per year. Economies of scale apply to CO$_2$ sequestration: the median sequestration site is shared among 3 RNG facilities, each transporting CO$_2$ an average of 26 miles. The amount of RNG produced makes up roughly 4\% of current natural gas usage in California, or roughly 3 times the current demand for utilizing natural gas as transportation fuel. 

The baseline utilizes municipal solid wastes (MSW) heavily: facilities annually process a total of 711,400 wet tons (wt) of food waste, 1.4 million wt of green waste, and 110,601 wt of grease, with an average transport distance of 20 miles. The scenario takes advantage of the colocation of agricultural activity and sequestration sites in the Central Valley---4.8 million wt of available crop residues and 23 million wt of manure are utilized. On average, crop residues are transported 26 miles and manure feedstocks are transported 15 miles from source to processing facility.

Our model generates networks of regional carbon negative biomethane production networks that exploit the diversity of feedstocks in the state in the optimal baseline scenario. Facilities tend to cluster near urban regions with readily available MSW. Broadly, we observe six distinct regions of agglomeration within California: in urban regions such as the greater Los Angeles area, Bay Area, Sacramento Valley; and agricultural regions around Bakersfield and Imperial County (Figure \ref{fig:baseline}). CO$_2$ sequestration in non-urban areas is economically feasible in all regions due to the widespread regional availability of sequestration sites. 

In the optimal scenario, we see significant heterogeneity in feedstock utilization in different parts of California and notable differences in landfill gas usage in urban areas. Landfill gas plays a significantly larger role in the Bay area and greater L.A. area compared to the Central Valley. In the Bay Area, landfill gas produces 87\% of the total RNG, compared to 6.5\% in Fresno. Overall, landfill gas make up roughly 39\% of the total RNG production. Manure, an important feedstock in certain regions, produces 70\% of the total RNG in Fresno and 99\% in Imperial County, taking advantage of the livestocks in these regions. The ability of the RNG-CCS system to process all varieties of biomass residues makes for an attractive option to produce carbon negative fuels across California. 

The baseline results above suggest that the current policy climate is sufficient to incentivize a robust build out of RNG-CCS systems in California. However, should existing policies be sharply curtailed, the RNG-CCS system could potentially lose its economic viability. To further investigate, we consider four policy scenarios: low and high policy support scenarios, a no RFS scenario, and no 45Q minimum threshold scenario (Table 1). In the "High Policy" scenario, we extend the policies to their allowed maximum. Specifically, we assume that the LCFS price is increased to \$200 per credit, RFS price to \$1.50 per credit, and 45Q remains at \$50 per ton CO$_2$. In the "Low Policy" scenario, we lower policies to their minimum: the LCFS price is decreased to \$20 per credit, the RFS credits become zero, while 45Q remains at \$50 per ton CO$_2$. The "No RFS" scenario entails removing RFS credit incentives while all else is as in the baseline scenario. We remove the minimum 100,000 CO$_2$ captured per year threshold to receive the 45Q tax credit in the "No 45Q Minimum Threshold" scenario while keeping all else equal. 

In the favorable policy environment of the "High Policy" scenario, 4 million tons of CO$_2$ is sequestered in 36 sites and 130 PJ of RNG is produced per year from 121 facilities. With significantly lower policy support in the "Low Policy" scenario, we still find that 1 million tons of CO$_2$ is sequestered in 8 sites and 32 PJ of RNG is produced per year from 37 facilities. Without RFS at baseline, we find that the baseline levels of LCFS and 45Q credits are sufficient to sustain a similar level of RNG-CCS build out in California. We still sequester 2.3 million tons of CO$_2$ in 14 sequestration sites and produce 73 PJ of RNG from 61 facilities (Figure \ref{fig:scenarios}). This represents a 22\% decrease in RNG production from the baseline scenario. Removing the minimum threshold 100,000 tons of captured CO$_2$ for 45Q eligibility does not increase the RNG-CCS build out in an optimal scenario. Instead, it produced 90 PJ of CH$_4$ in 95 facilities and sequesters 2.8 million tons of CO$_2$ in 27 sequestration sites. 

Profits range from \$3 and \$32/GJ under a low and high policy support environment, respectively (Figure \ref{fig:cost_revenue}). Without RFS, we find that profits only decrease slightly compared to the baseline from \$11/GJ to \$7.4/GJ. Profits remain roughly similar when we remove the minimum 45Q threshold, earning an average of \$11.50/GJ.  LCFS credits remain a significant source of revenue across all scenarios. Especially without RFS at baseline, LCFS make up an overwhelming majority (81\%) of the revenues earned. Surprisingly, CCS-related costs are small relative to biomass processing related costs across all scenarios, making up roughly 10\% of the total baseline cost. Biomass processing related costs dominate: levelized digester costs account for nearly 45\% (\$6.50/GJ) of the total baseline cost, and feedstock transport costs account for 42\% (\$5.93/GJ) of the total baseline cost. Across varying levels of policy support, we find that the RNG-CCS system is still able to profitably produce carbon-negative transportation fuel. 

Carbon removal through renewable natural gas production with CCS in California is cost-effective and profitable across the range of scenarios we explore. This niche market for BECCS holds numerous implications for California’s climate mitigation efforts, waste management, and energy policy. We discuss the impacts of existing policies on the build-out of RNG-CCS system in California (Figure \ref{fig:sensitivity}). 

We find that federal and state support for carbon sequestration is a large driver of near-term BECCS deployment. To examine the marginal impact of various policies, we perform sensitivity analysis parametrically. In the baseline case, existing policies are capable of incentivizing a robust network of RNG-CCS systems in California while generating \$11/GJ of profits. We find that CO$_2$ sequestration and RNG production are most sensitive to LCFS credit prices relative to other policy drivers. All else equal, when LCFS credits are reduced from \$100/tCO$_2$ to \$0/tCO$_2$, we sequester 1.1 million tons of CO$_2$, compared to 2.9 million tons of CO$_2$ at the baseline. We only produce 37 PJ of RNG compared to 93 PJ at the baseline. When LCFS credits are \$200/tCO$_2$, we sequester over 3.8 million tons of CO$_2$ and produce 123 PJ of RNG. RFS also appears to be an efficient policy driver to incentivize carbon negative systems relative to the LCFS. CO$_2$ sequestered varies from 2.6 to 4 MtCO$_2$/yr and RNG production ranges 89 and 131 PJ per year while incentivizing profits upwards of \$19/GJ when we vary the RFS credit price. Taken together, the results suggest that LCFS and RFS are necessary drivers of RNG-CCS systems in California. While the LCFS is only officially extended through 2030, California governor Newsom issued an executive order to the Air Resources Board (ARB) to extend the LCFS beyond 2030 \cite{execorder}.

Moreover, the 45Q tax credit also plays a complimentary role to LCFS and RFS credits. As it is currently designed, the tax credit has a minimum CO$_2$ captured threshold of 100,000 tons. In popular press, many argue that the minimum threshold should be removed for less project finance risks \cite{burns_jacobson_2021}. We show that the threshold is irrelevant in the context of RNG-CCS where the threshold acts as an agglomeration force without dampening RNG production. However, the threshold might act as a significant obstacle in pilot projects with relatively newer technologies or do not produce transport fuels \cite{mcqueen2020cost}.

California is well-suited for systems such as the RNG-CCS system presented here. The feedstock heterogeneity across regions and abundance of potential of sequestration sites throughout the state create rich opportunities for low-cost and near-term carbon negative systems. Moreover, fuel production with CCS could be an important part of the energy supply mix that satisfies low-carbon fuel mandates in California and the United States. Other low-carbon and carbon-negative fuel technologies with CCS such as ethanol coupled with CCS, enhanced oil recovery (EOR), Direct Air Capture with EOR (DAC-EOR), and many others, could become important under current policy conditions (\citenum{sanchez_near-term_2018}, \citenum{sanchez_literature_2020}). 

Yet, we stress that RNG-CCS is not a substitute for electrification in California—at baseline, it produces only 4\% of current natural gas demand. If we electrify rapidly according to the E3 PATHWAYS high building electrification scenario, RNG-CCS at baseline would meet 42\% of the RNG demand and 14\% of the overall natural gas demand \cite{mahone_deep_nodate}. In contrast, if we do not electrify effectively, the baseline RNG production would meet over 100\% of the RNG demand and only 7\% of the overall natural gas demand. Despite this, the RNG-CCS system could be attractive to supply transportation fuel in California, and serve as a first market for CCS in California.

RNG-CCS can promote broader CCS deployment in California. We find that sharing sequestration sites across different CCS projects can further lower the capital cost of CO$_2$ sequestration while ensuring effective monitoring and verification. Existing efforts by the Department of Energy’s C2SAFE program aim to sequester 50 million tons of CO$_2$ over 20 to 30 years \cite{trautz_california_2018}. Furthermore, sequestration and abatement credits to drive CCS first markets can develop experience, project financing, policies, and business models for CCS. Within California, the Central Valley is prime to emerge as a hub for CCS. \citet{baker_getting_2020} describe the central valley as “ready to go” to address CO$_2$ storage needs. Further work is necessary to enable sequestration at sites near urban areas, which here we assume cannot be utilized. 

We note several key limitations to our current modeling framework. First, we assume that the natural gas produced can be injected into the natural gas pipeline network in California. However, depending on feedstock input and upgrading technologies, further RNG upgrading to remove any trace amounts of N$_2$, H$_2$, H$_2$S, O$_2$, and siloxanes may be required. Further, many existing natural gas pipelines lack adequate capacity \cite{gas_technology_institue_low_nodate}. We also consider only one type of waste management in California: anaerobic digestion. Composting is  another viable waste management policy. We do not evaluate the tradeoffs between composting biomass residues and using biomass residues to produce RNG via anaerobic digestion. We assume that any potential CH$_4$ emissions from leakage in the natural gas system is negligible to the climate impacts of reduced GHG intensity of the RNG and sequestered CO$_2$. However, \citet{grubert_at_2020} finds that RNG systems could be climate intensive depending on the methane feedstock. The analysis in this paper is on a yearly basis, but biomass residues, especially agricultural residues, are seasonal. That is, while average anaerobic digester capacity predicted by the model is relatively smaller, in reality, digesters need to be outsized for intermittent large inflows while sitting idle for some time. Lastly, we consider a global optimization. That is, we optimize for one collective objective function instead of individual firm objective functions. Firms make decisions individually, not globally, and a facility-level optimization might yield different results. The significant economies of scale involved obviate a need for cost-sharing or redistribution to ensure the profitability and impact of an RNG-CCS system at a facility scale.  

Profitable RNG production with CCS potentially catalyzes markets and technologies for CO$_2$ capture, transport, and removal in California. In particular, RNG with CCS can address both the waste management and climate mitigation needs in California. Producing up to 93 PJ of RNG and sequestering up to 2.9 million tons of CO$_2$ per year, RNG-CCS can stimulate a robust carbon-negative system while providing a profitable first market for BECCS. The deployment of RNG-CCS systems in California can accelerate efforts for BECCCS internationally and in California. As California considers a broad range of carbon removal strategies to meet its goal of net neutrality economy-wide in 2045, we suggest that RNG-CCS could play an early and prominent role. More broadly, RNG-CCS' success in California can lead to further innovation and investment in BECCS beyond state borders and substantially impact the federal climate policy landscape.

\section*{Acknowledgements}
We thank Corrine Scown, Curtis Oldenburg, Greg Kester, Hanna Breunig, Maureen Walsh, Preston Jordan, Tom Richard, and Senorpe Asem-Hiablie for especially helpful discussions. We acknowledge the support of the Conservation 2.0 program.

\newpage

\begin{table}[!htbp]
    \centering
    \caption{Policy Scenarios}
    \begin{tabular}{lccccc}
        \hline \hline 
         Policies (units) & Baseline & No RFS & No 45Q Threshold & High Policy & Low Policy  \\
          & & & &  \\
         LCFS (\$/tCO$_2$) & \$100 & \$100 & \$100 & \$200 & \$20 \\ 
         RFS (\$/GGE) & \$0.25 & \$0 & \$0.25 & \$1.50 & \$0 \\
         45Q (\$/tCO$_2$) & \$50 &  \$50 & \$50 & \$50 & \$50 \\
         45Q Minimum (tCO$_2$/year) & 100,000 & 100,000 & 0 & 100,000 & 100,000 \\
         \hline \hline
    \end{tabular}
    \label{tab:my_label}
\end{table}

\newpage

\newcommand{\subsubfloat}[2]{%
  \begin{tabular}{@{}c@{}}#1\\#2\end{tabular}%
}

\vspace{0.3cm}

\begin{singlespace}
\begin{footnotesize}
Available biomass residues (green), existing anaerobic digesters and landfills (blue), and potential CO2 sequestration sites (grey) in California. Colocation of residues, existing digesters, and potential CO$_2$ sequestration sites across California reduces biomass and CO$_2$ transportation costs.
\end{footnotesize}
\end{singlespace}

\newpage

\begin{singlespace}
\begin{footnotesize}
(a) Process flow diagram and (b) Model flow chart for RNG-CCS systems. Panel A shows major processes in the RNG-CCS system. We consider relevant system boundaries for lifecycle CO$_2$ emissions assessment consistent with California’s Low Carbon Fuels Standard. We consider CO$_2$ emissions from CO2 compression, transportation, and sequestration. Panel B shows major features of our optimization model. Our optimization maximizes total profits, including binary decision variables to activate a particular facility or sequestration site, alongside continuous production decision variables. 
\end{footnotesize}
\end{singlespace}

\newpage

\newpage

\setcounter{figure}{2}   

\begin{figure}[H]
    \begin{subfigure}{0.5\textwidth}
    \addtocounter{subfigure}{3} 
    \caption{Fresno Area}
        \centering
    \includegraphics[width=\linewidth]{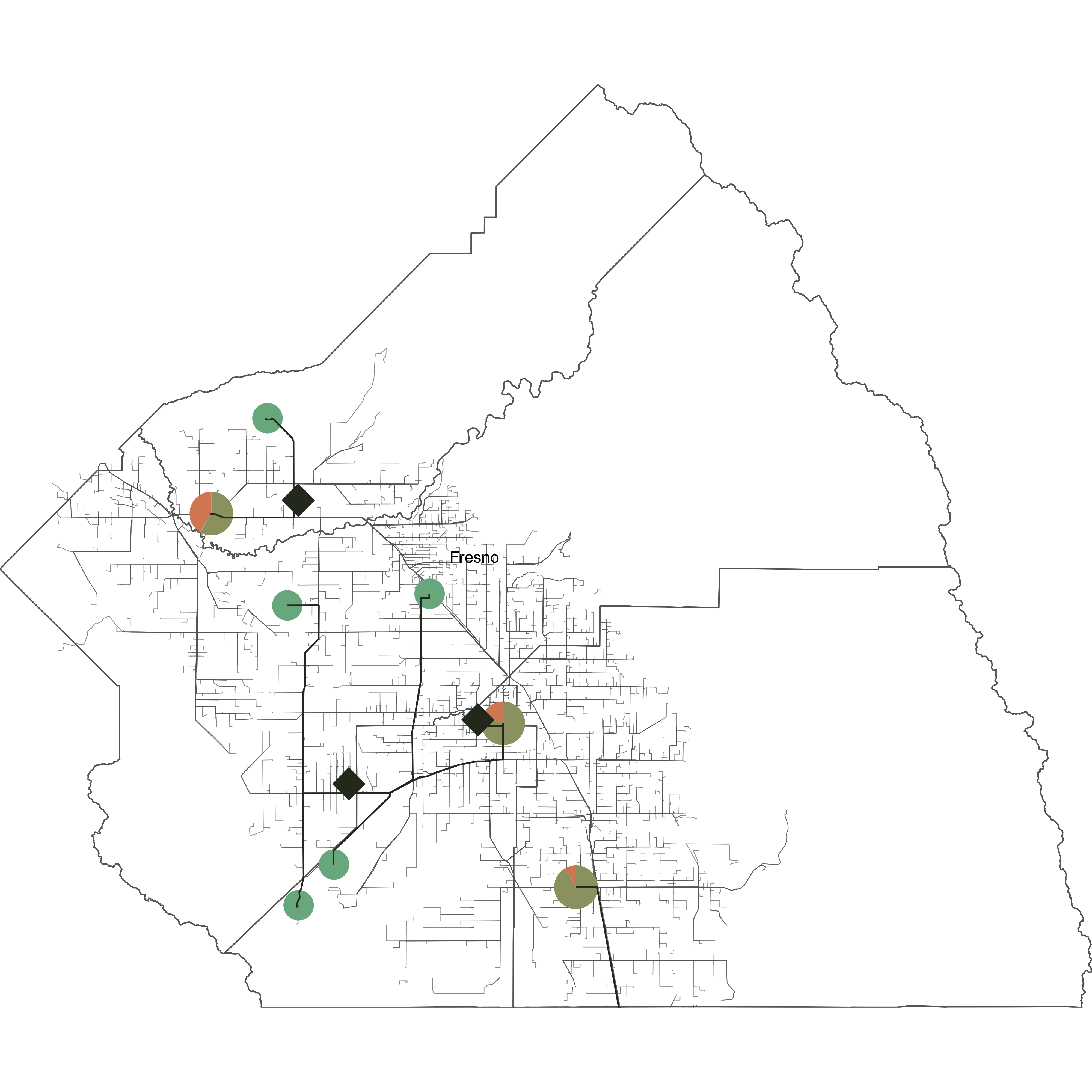}
\end{subfigure}%
\begin{subfigure}{0.5\textwidth}
    \caption{Imperial County}
        \centering
    \includegraphics[width=0.8\linewidth]{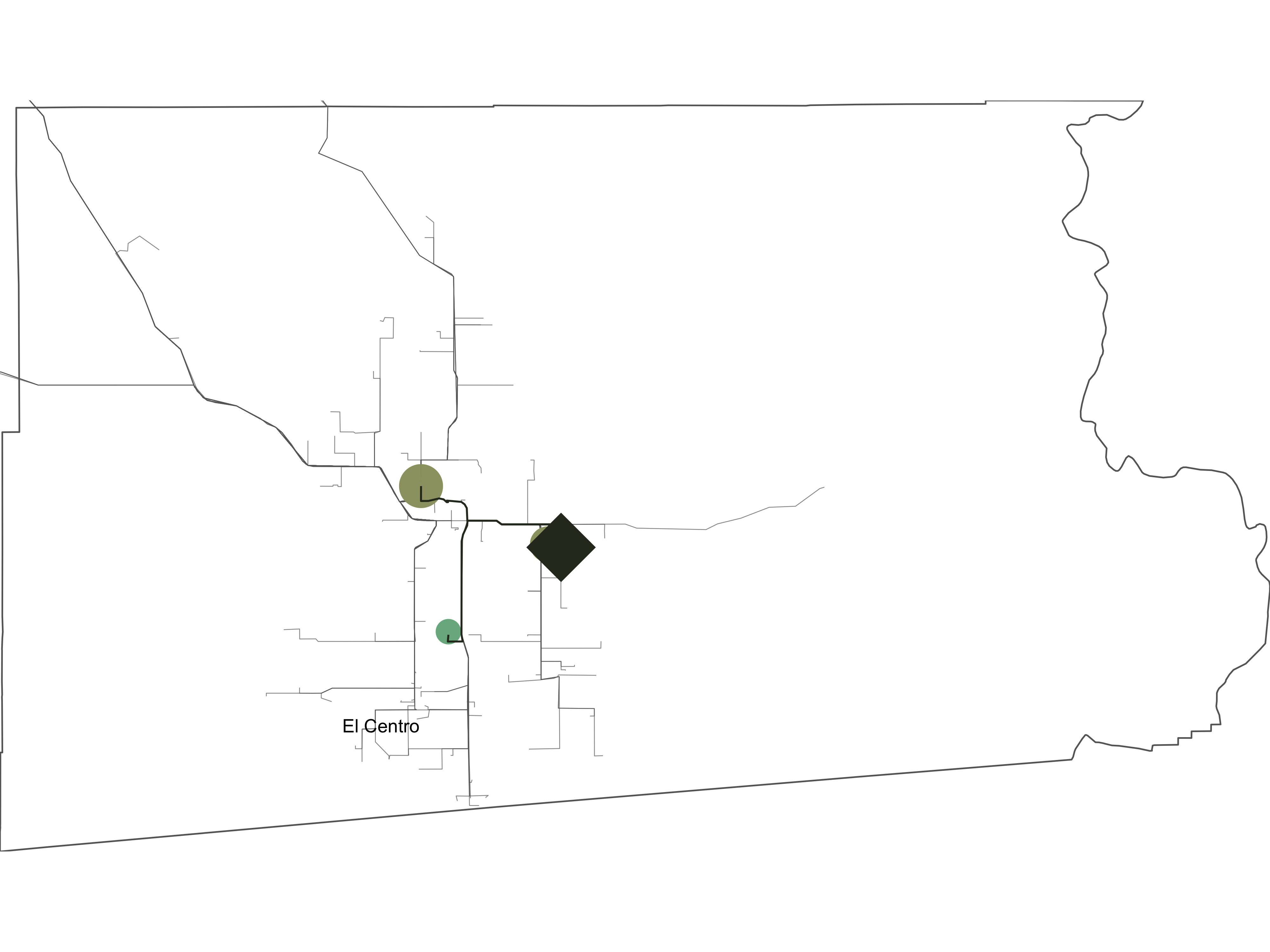}
\end{subfigure}

    \begin{subfigure}{0.5\textwidth}
        \caption{Southern California}
            \centering
    \includegraphics[width=0.8\linewidth]{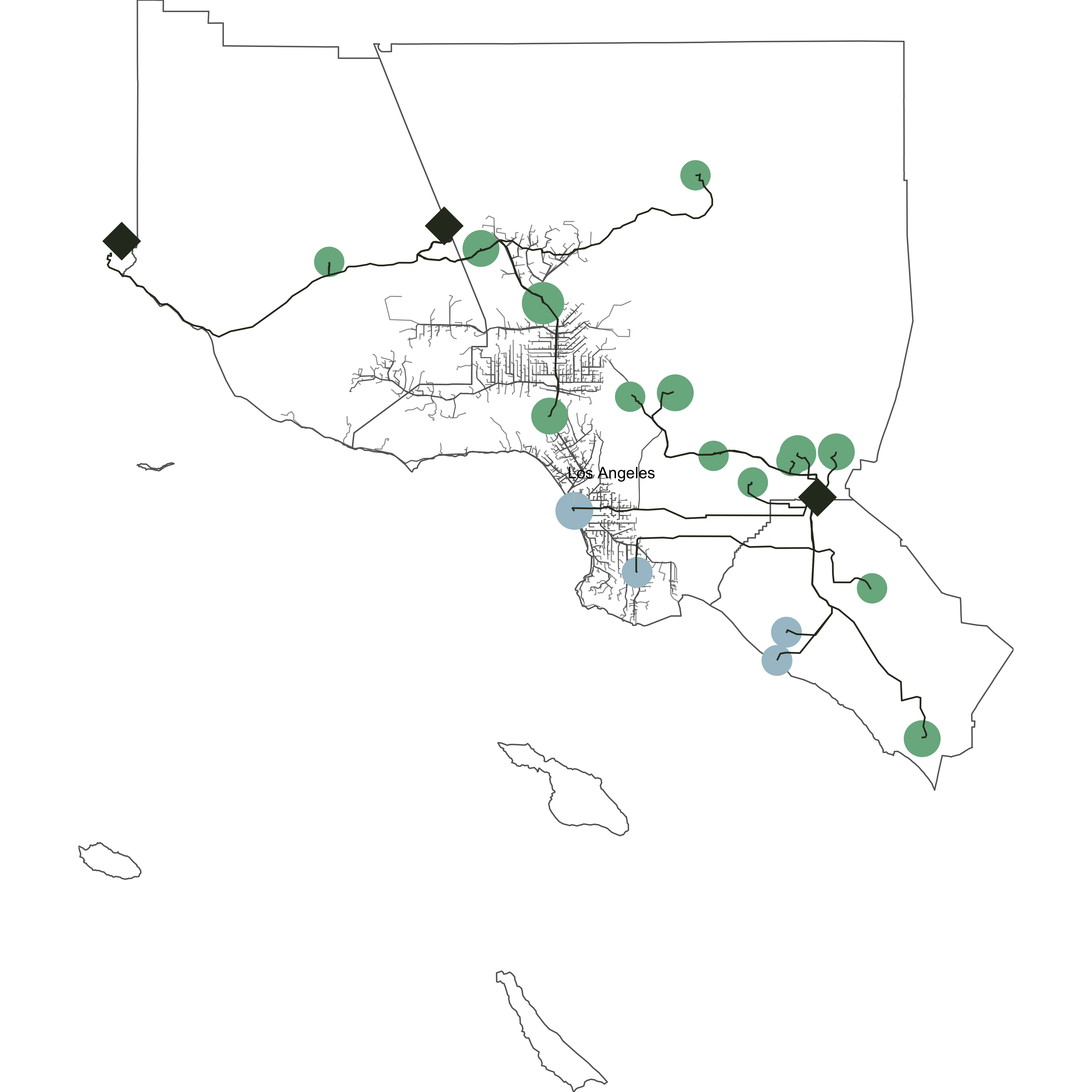}

\end{subfigure}%
\begin{subfigure}{0.5\textwidth}
    \caption{Bay Area}
        \centering
    \includegraphics[width=0.8\linewidth]{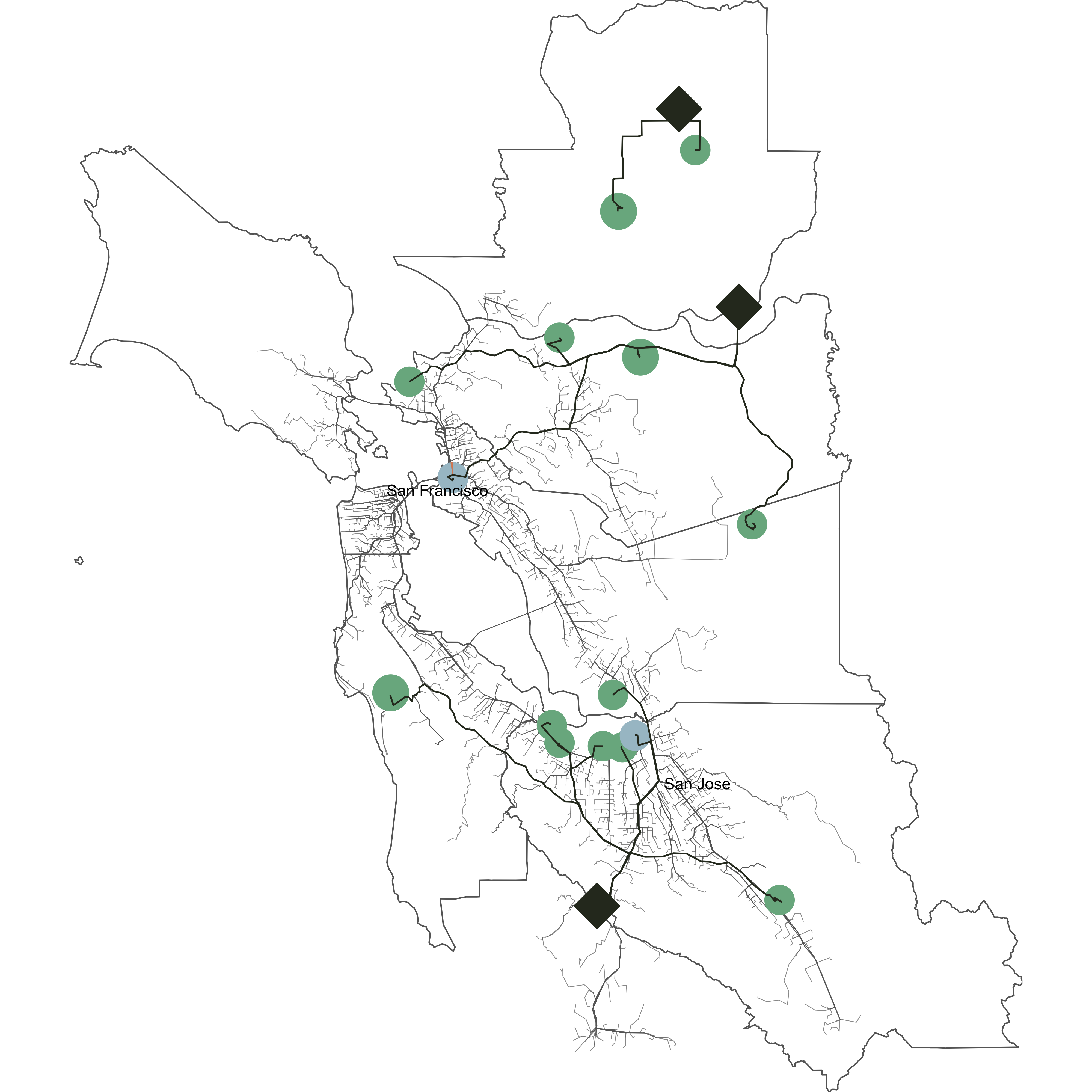}

\end{subfigure}
\end{figure}

\begin{singlespace}
\begin{footnotesize}
Baseline RNG-CCS system in (a) California and (b-g) in prominent regional agglomerations. Across California, RNG-CCS systems produce 119 PJ of RNG from 85 facilities per year, and sequester 4.1 million tCO$_2$ in 19 sequestration sites per year. In panels B through G, we focus on the emerging regional agglomerations. Extensive biomass residue and CO$_2$ sequestration collection networks are formed, enabled by small-volume truck transportation.
\end{footnotesize}
\end{singlespace}

\newpage

\begin{figure}[H]
\centering
\caption{Policy Scenarios}

\begin{subfigure}[c]{\textwidth}
        \caption{Baseline}
        \centering
        \includegraphics[width=0.8\linewidth]{Figures/policy_california.png}
\end{subfigure}

\begin{subfigure}[c]{\textwidth}
        \caption{No RFS}
        \centering
        \includegraphics[width=0.8\linewidth]{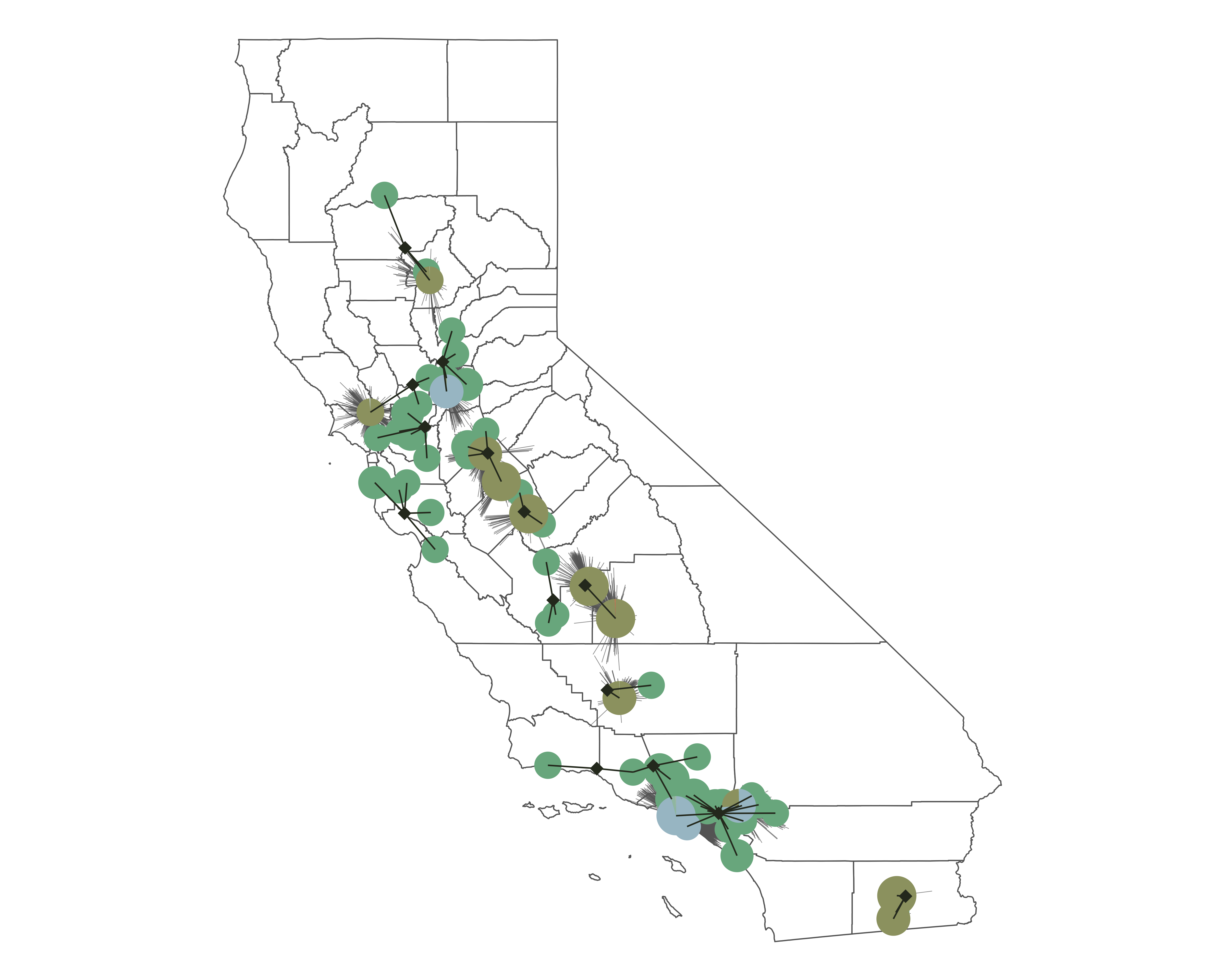}
\end{subfigure} 
\label{fig:scenarios}
\end{figure}
  
\setcounter{figure}{3} 

\begin{figure}[H]
\centering
\begin{subfigure}[c]{\textwidth}
    \addtocounter{subfigure}{2} 
    \caption{Low Policy}
    \centering
    \includegraphics[width=0.8\linewidth]{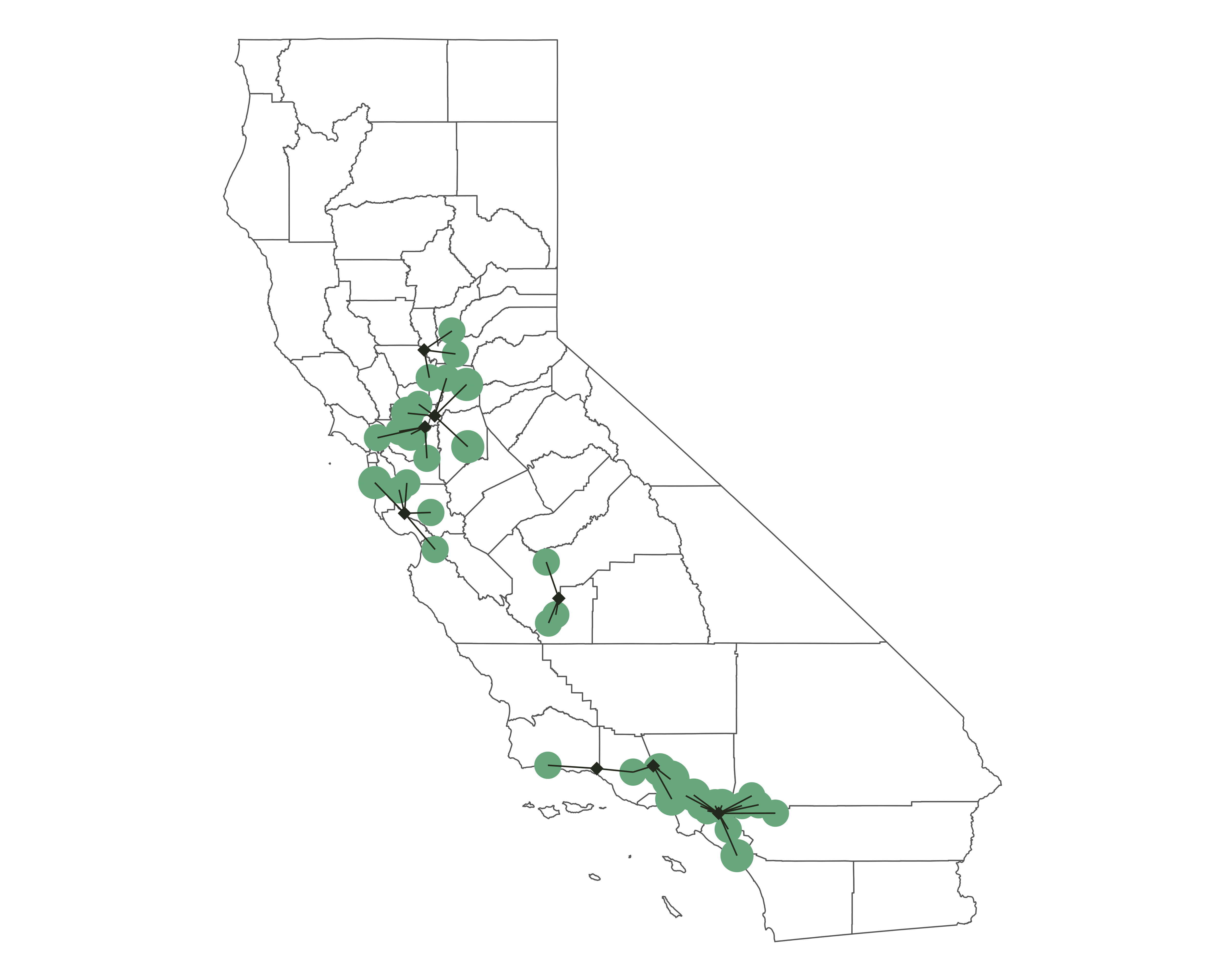}
\end{subfigure}

\begin{subfigure}[c]{\textwidth}
    \caption{High Policy}
    \centering
    \includegraphics[width=0.8\linewidth]{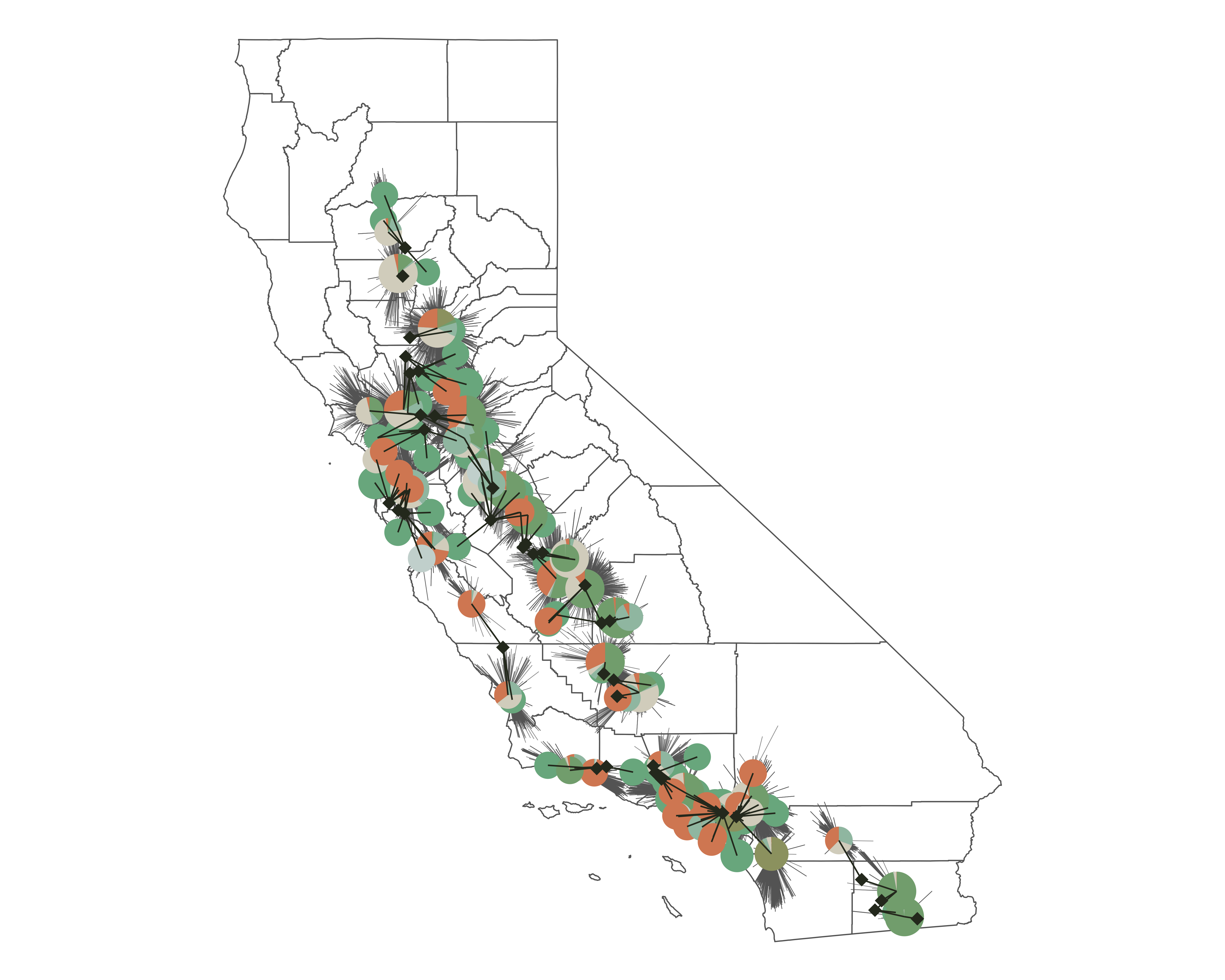}
\end{subfigure}
\end{figure}

\setcounter{figure}{3} 

\begin{figure}[H]
\begin{subfigure}[c]{\textwidth}
    \addtocounter{subfigure}{4} 
    \caption{No 45Q Threshold}
    \centering
    \includegraphics[width=0.8\linewidth]{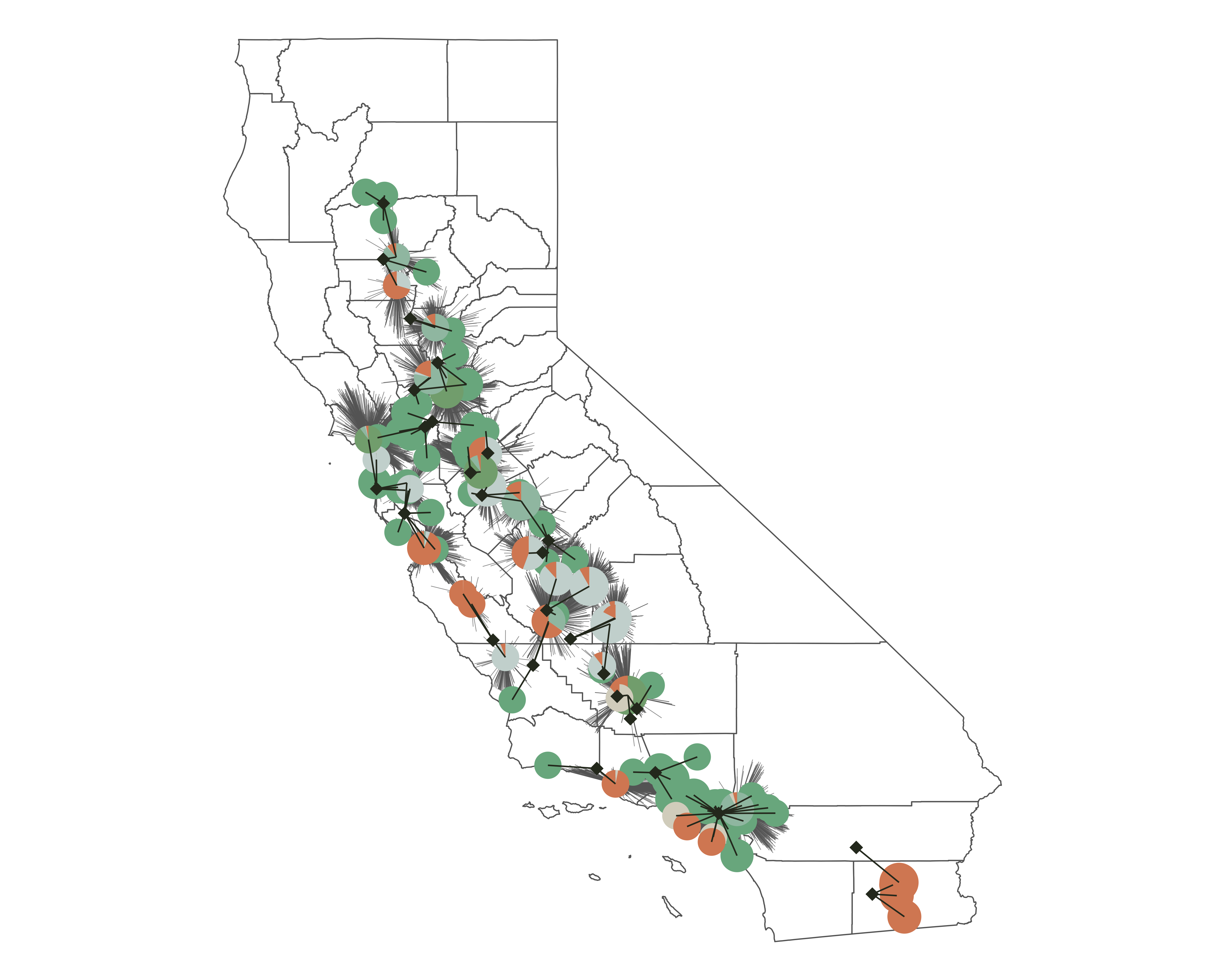}
\end{subfigure}
\end{figure}
\begin{singlespace}
\begin{footnotesize}
Statewide RNG-CCS system for various policy scenarios: (a) Baseline (b) No RFS, (c) Low Policy, (d) High Policy, and (e) No 45Q Minimum Threshold. RNG-CCS systems are robust to varying levels of policy support.
\end{footnotesize}
\end{singlespace}

\newpage

\begin{figure}[H]
    \centering
    \caption{Cost and Revenues}
    \includegraphics[width=\textwidth]{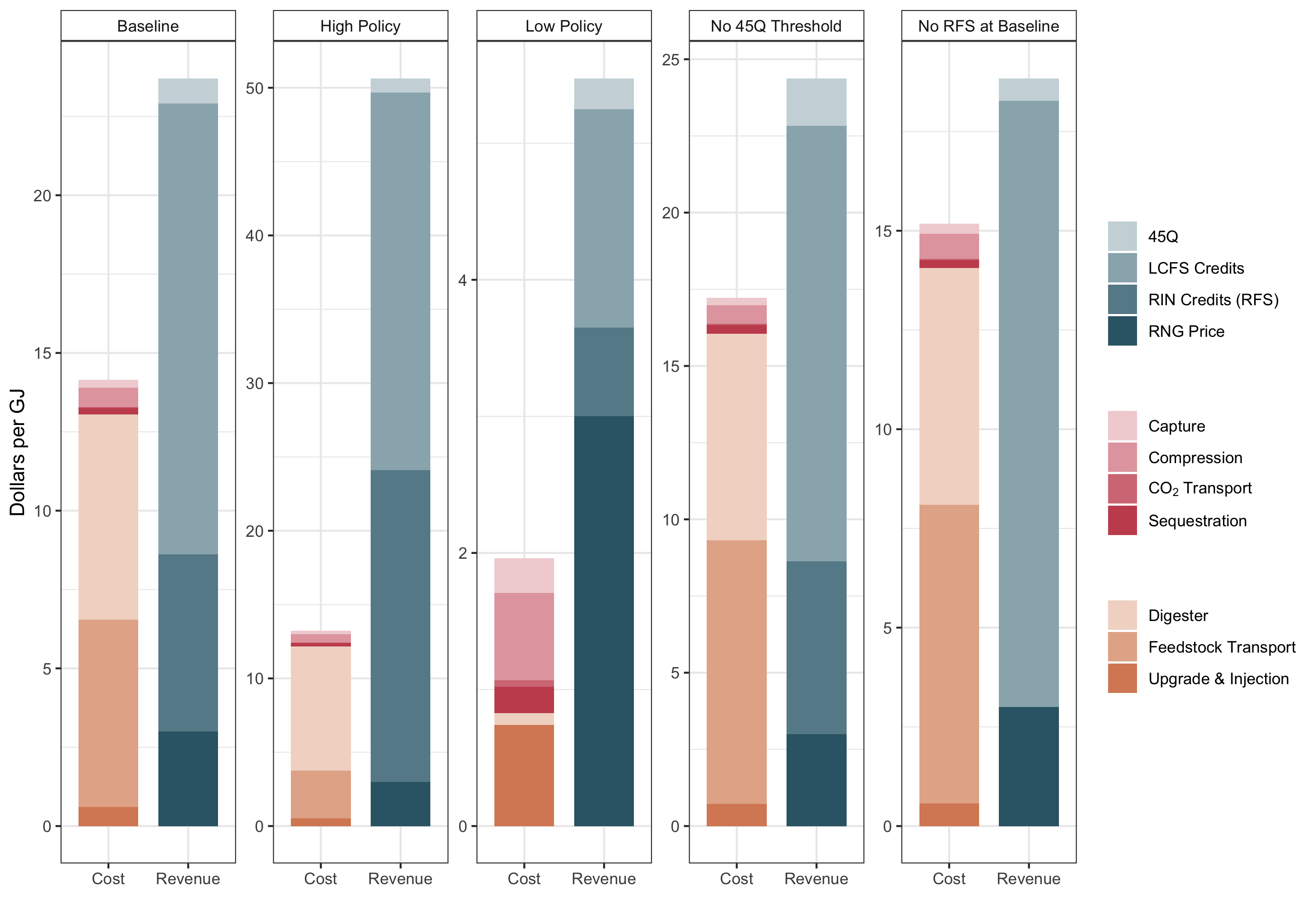}
    \label{fig:cost_revenue}
\end{figure}

\begin{singlespace}
\begin{footnotesize}
Costs and revenues (\$/ GJ) for RNG-CCS system for various policy scenarios. Costs are separated into two technology categories: CCS-related (red) and biomass processing-related (orange). Across all scenarios, CCS-related costs are a small fraction of total costs. Revenues from the LCFS make up a large share of revenue across all scenarios.
\end{footnotesize}
\end{singlespace}

\newpage

\begin{figure}[H]
    \centering
    \caption{Sensitivity analysis}
    \begin{subfigure}{\textwidth}
        \centering
        \caption{CH$_4$ production per year}
        \includegraphics[width=0.75\linewidth]{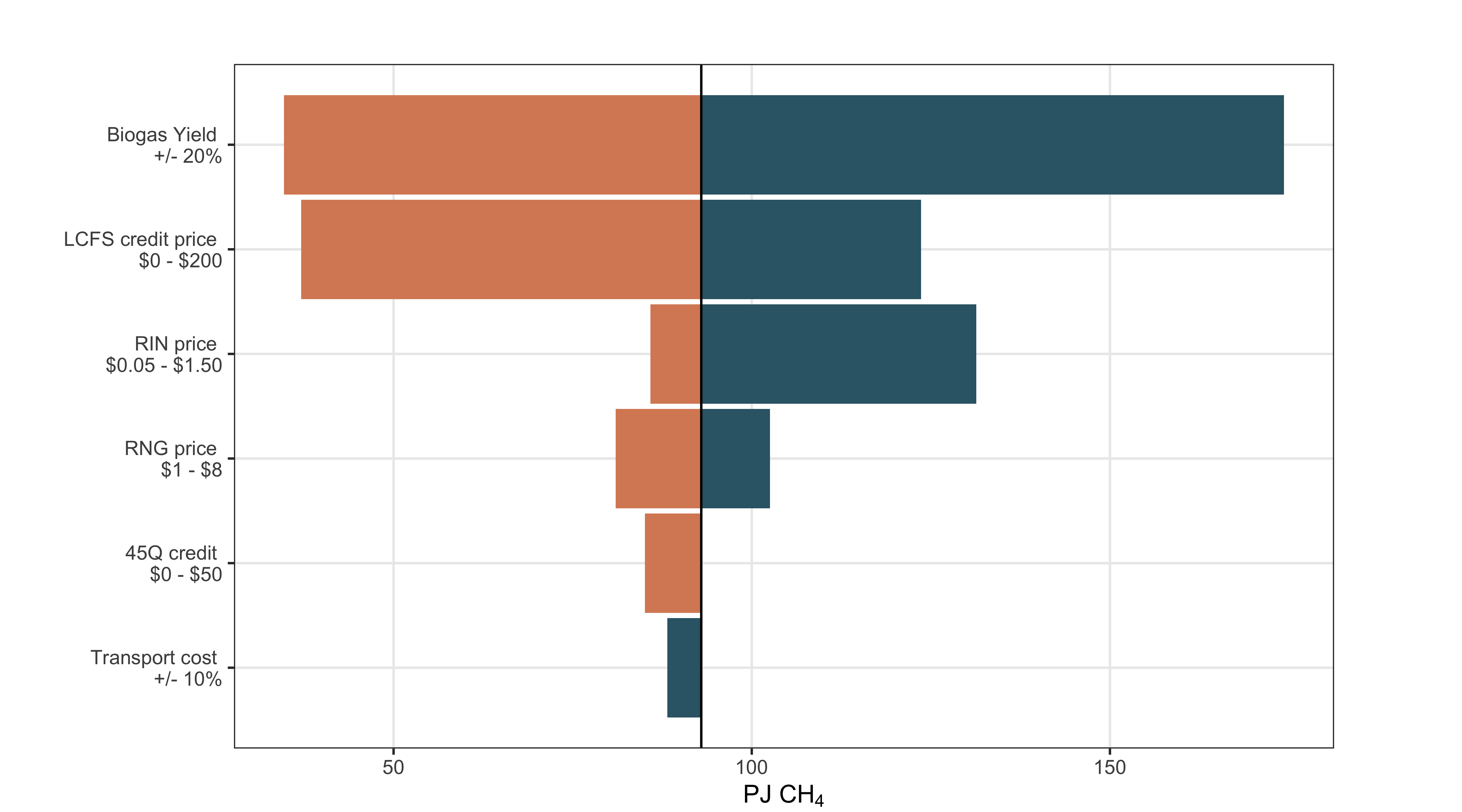}
    \end{subfigure}
    
    \begin{subfigure}{\textwidth}
        \centering
        \caption{CO$_2$ sequestration per year}
        \includegraphics[width=0.75\linewidth]{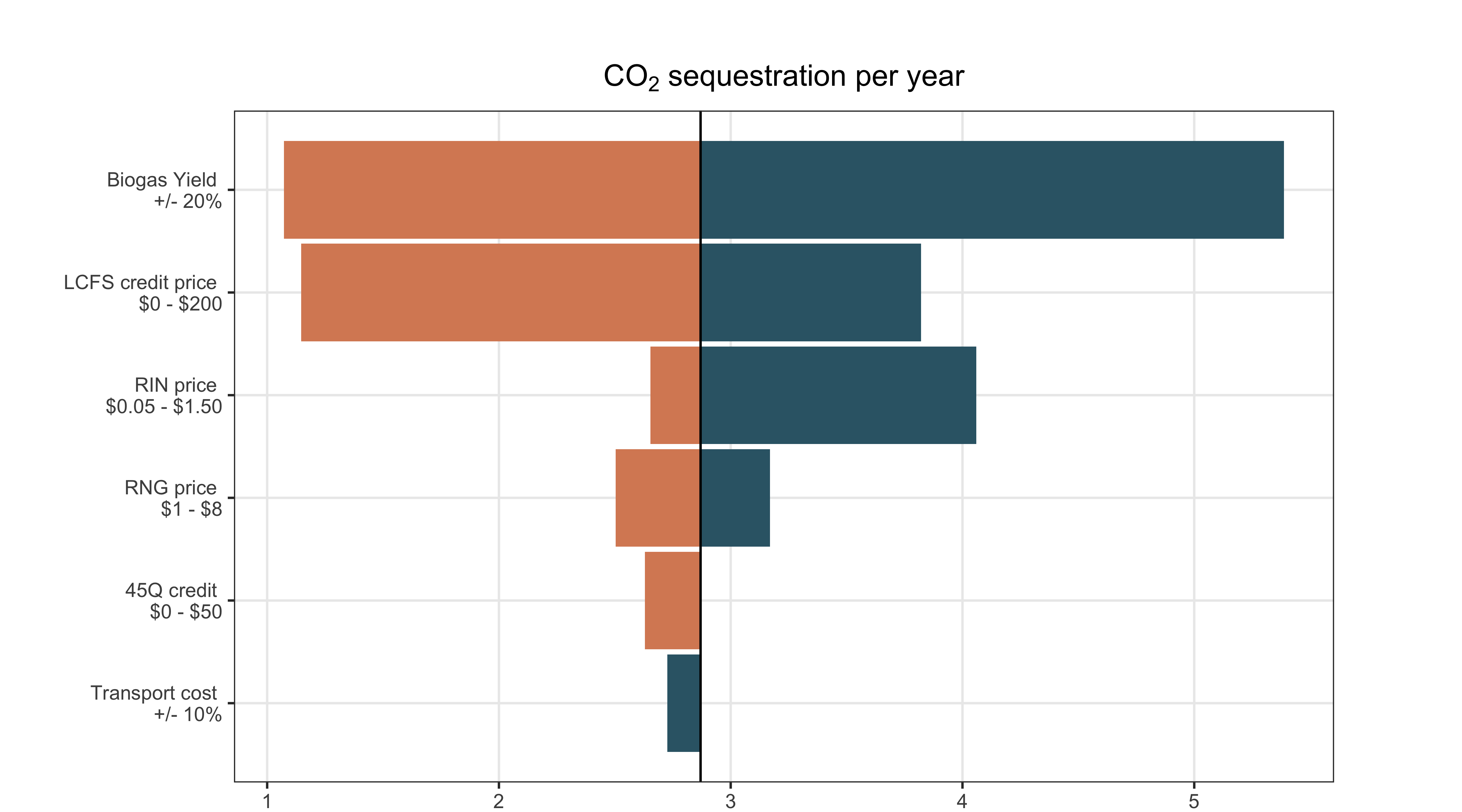}
    \end{subfigure}
    
    \begin{subfigure}{\textwidth}
        \centering
        \caption{Profits per year}
        \includegraphics[width=0.75\linewidth]{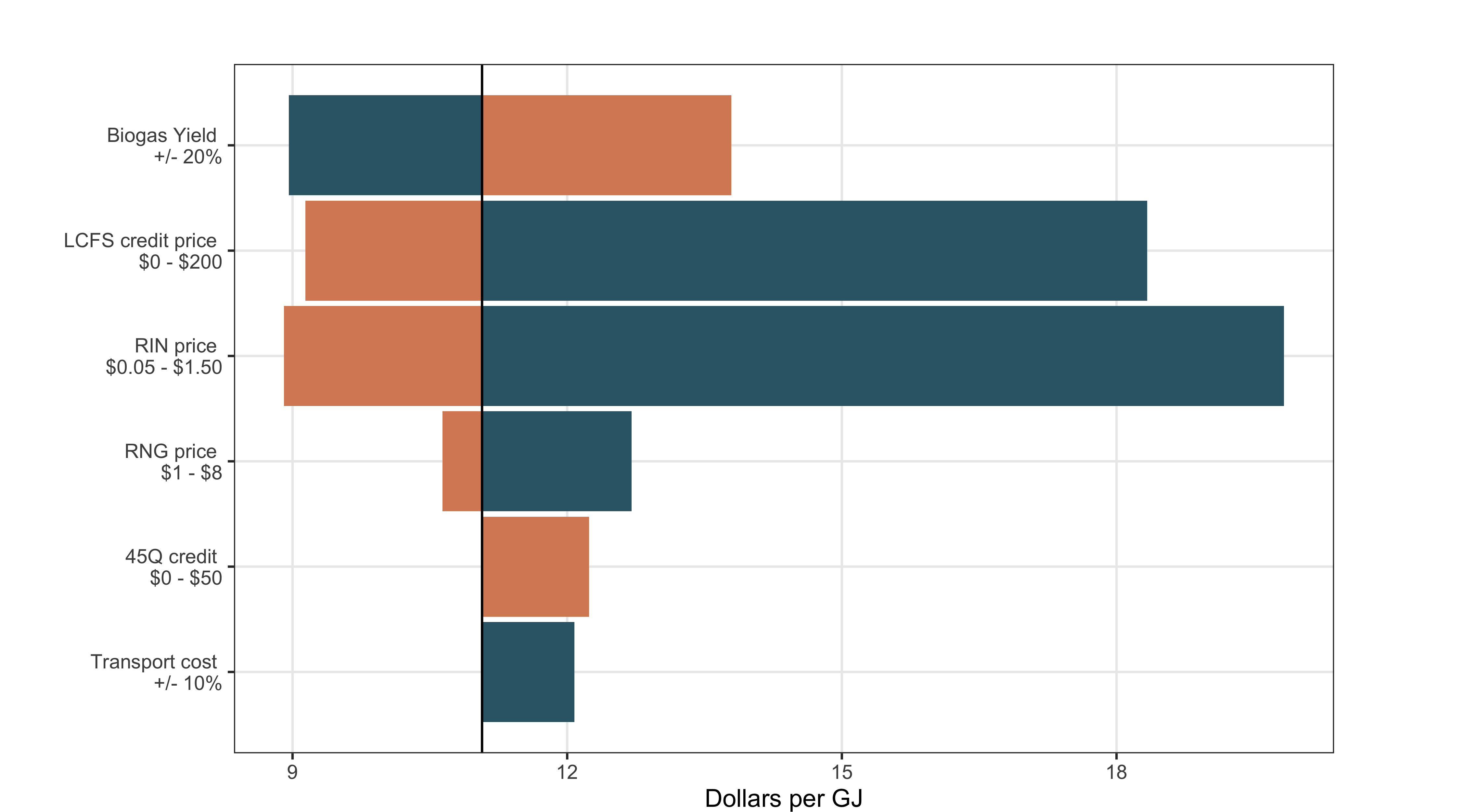}
    \end{subfigure}

    \label{fig:sensitivity}
\end{figure}

\begin{singlespace}
\begin{footnotesize}
Parametric sensitivity analysis of yearly (a) CH4 production (b) CO$_2$ sequestration, and (c) profit. We vary biogas yield, LCFS price, RIN credit price, RNG price, 45Q credits, and transport cost. Average profits are rarely fall below the baseline \$12/GJ, but can reach as high as \$37/GJ when RFS prices are at their highest.
\end{footnotesize}
\end{singlespace}

\newpage
\begin{singlespace}
\bibliography{biogas_ccs}

\begin{thebibliography}{37}
\providecommand{\natexlab}[1]{#1}
\providecommand{\url}[1]{\texttt{#1}}
\expandafter\ifx\csname urlstyle\endcsname\relax
  \providecommand{\doi}[1]{doi: #1}\else
  \providecommand{\doi}{doi: \begingroup \urlstyle{rm}\Url}\fi

\bibitem[Aines et~al.()Aines, Amador, Babson, Bennett, Cave, Chen, Dipple,
  Field, Filley, Friedmann, Kelemen, Jacobson, Funk, Lehmann, Levy, Lucas,
  Matchett, {McCarty}, Mooney, Neale, Park, Pett-Ridge, Powell, Sanchez, Song,
  Stechel, Sun, Sun, Swan, Wilcox, Woodbury, and Zelikova]{aines_building_2019}
Roger Aines, Giana Amador, David Babson, Drew Bennett, Etosha Cave, Wei Chen,
  Gregory Dipple, John~L. Field, Tim Filley, S.~Julio Friedmann, Peter Kelemen,
  Rory Jacobson, Jason Funk, Johannes Lehmann, Charlotte Levy, Matt Lucas,
  Karin Matchett, Jessica~L. {McCarty}, Sian Mooney, Nathan~R. Neale,
  Ah-Hyung~Alissa Park, Jennifer Pett-Ridge, Joe Powell, Daniel~L. Sanchez,
  Chunshan Song, Ellen Stechel, Nannan Sun, Yuhan Sun, Amy Swan, Jennifer
  Wilcox, Peter~B. Woodbury, and Jane Zelikova.
\newblock Building a new carbon economy: An innovation plan.
\newblock URL
  \url{https://static1.squarespace.com/static/5b9362d89d5abb8c51d474f8/t/5b98383aaa4a998909c4b606/1536702527136/ccr02.innovationplan.FNL.pdf}.

\bibitem[Sanchez et~al.({\natexlab{a}})Sanchez, Johnson, {McCoy}, Turner, and
  Mach]{sanchez_near-term_2018}
Daniel~L. Sanchez, Nils Johnson, Sean~T. {McCoy}, Peter~A. Turner, and
  Katharine~J. Mach.
\newblock Near-term deployment of carbon capture and sequestration from
  biorefineries in the united states.
\newblock 115\penalty0 (19):\penalty0 4875--4880, {\natexlab{a}}.
\newblock ISSN 0027-8424, 1091-6490.
\newblock \doi{10.1073/pnas.1719695115}.
\newblock URL \url{https://www.pnas.org/content/115/19/4875}.
\newblock Publisher: National Academy of Sciences Section: Physical Sciences.

\bibitem[Anderson and Peters()]{anderson_trouble_2016}
Kevin Anderson and Glen Peters.
\newblock The trouble with negative emissions.
\newblock 354\penalty0 (6309):\penalty0 182--183.
\newblock ISSN 0036-8075, 1095-9203.
\newblock \doi{10.1126/science.aah4567}.
\newblock URL \url{https://science.sciencemag.org/content/354/6309/182}.
\newblock Publisher: American Association for the Advancement of Science
  Section: Perspective.

\bibitem[Kearns()]{kearns_waste--energy_2019}
David~T Kearns.
\newblock Waste-to-energy with {CCS}: A pathway to carbon-negative power
  generation.
\newblock page~11.

\bibitem[Möllersten et~al.()Möllersten, Yan, and
  R.~Moreira]{mollersten_potential_2003}
Kenneth Möllersten, Jinyue Yan, and Jose R.~Moreira.
\newblock Potential market niches for biomass energy with {CO}2 capture and
  storage—opportunities for energy supply with negative {CO}2 emissions.
\newblock 25\penalty0 (3):\penalty0 273--285.
\newblock ISSN 0961-9534.
\newblock \doi{10.1016/S0961-9534(03)00013-8}.
\newblock URL
  \url{http://www.sciencedirect.com/science/article/pii/S0961953403000138}.

\bibitem[Bellamy and Geden()]{bellamy_govern_2019}
Rob Bellamy and Oliver Geden.
\newblock Govern {CO} 2 removal from the ground up.
\newblock 12\penalty0 (11):\penalty0 874--876.
\newblock ISSN 1752-0908.
\newblock \doi{10.1038/s41561-019-0475-7}.
\newblock URL \url{https://www.nature.com/articles/s41561-019-0475-7}.
\newblock Number: 11 Publisher: Nature Publishing Group.

\bibitem[Nemet et~al.()Nemet, Callaghan, Creutzig, Fuss, Hartmann, Hilaire,
  Lamb, Minx, Rogers, and Smith]{nemet_negative_2018}
Gregory~F. Nemet, Max~W. Callaghan, Felix Creutzig, Sabine Fuss, Jens Hartmann,
  Jérôme Hilaire, William~F. Lamb, Jan~C. Minx, Sophia Rogers, and Pete
  Smith.
\newblock Negative emissions—part 3: Innovation and upscaling.
\newblock 13\penalty0 (6):\penalty0 063003.
\newblock ISSN 1748-9326.
\newblock \doi{10.1088/1748-9326/aabff4}.
\newblock URL
  \url{https://iopscience.iop.org/article/10.1088/1748-9326/aabff4/meta}.
\newblock Publisher: {IOP} Publishing.

\bibitem[Breunig et~al.()Breunig, Huntington, Jin, Robinson, and
  Scown]{breunig_temporal_2018}
Hanna~Marie Breunig, Tyler Huntington, Ling Jin, Alastair Robinson, and
  Corinne~Donahue Scown.
\newblock Temporal and geographic drivers of biomass residues in california.
\newblock 139:\penalty0 287--297.
\newblock ISSN 0921-3449.
\newblock \doi{10.1016/j.resconrec.2018.08.022}.
\newblock URL
  \url{http://www.sciencedirect.com/science/article/pii/S0921344918303148}.

\bibitem[Baker et~al.()Baker, Stolaroff, Peridas, Pang, Goldstein, Lucci, Li,
  Slessarev, Pett-Ridge, Ryerson, Wagoner, Kirkendall, Aines, Sanchez, Cabiyo,
  Baker, {McCoy}, Uden, Runnebaum, Wilcox, and {McCormick}]{baker_getting_2020}
Sarah~E. Baker, Joshsuah~K. Stolaroff, George Peridas, Simon~H. Pang, Hannah~M.
  Goldstein, Felicia~R. Lucci, Wenqin Li, Eric~W. Slessarev, Jennifer
  Pett-Ridge, Frederick~J. Ryerson, Jeff~L. Wagoner, Whitney Kirkendall,
  Roger~D. Aines, Daniel~L. Sanchez, Bodie Cabiyo, Joffre Baker, Sean {McCoy},
  Sam Uden, Ron Runnebaum, Jennifer Wilcox, and Colin {McCormick}.
\newblock Getting to neutral: Options for negative emissions in california.
\newblock URL
  \url{https://www-gs.llnl.gov/content/assets/docs/energy/Getting_to_Neutral.pdf}.

\bibitem[Witcover()]{witcover_status_2018}
Julie Witcover.
\newblock Status review of california’s low carbon fuel standard, 2011–2018
  q1 september 2018 issue.
\newblock URL \url{https://escholarship.org/uc/item/445815cd}.

\bibitem[Board()]{california_air_resources_board_low_2019}
California Air~Resources Board.
\newblock Low carbon fuel standard amendments.
\newblock URL \url{https://ww2.arb.ca.gov/rulemaking/2019/lcfs2019}.

\bibitem[lar()]{lara_bill_nodate}
Bill text - {SB}-1383 short-lived climate pollutants: methane emissions: dairy
  and livestock: organic waste: landfills.
\newblock URL
  \url{https://leginfo.legislature.ca.gov/faces/billNavClient.xhtml?bill_id=201520160SB1383}.

\bibitem[{US}~{EPA}({\natexlab{a}})]{us_epa_agstar_2014}
{OAR} {US}~{EPA}.
\newblock {AgSTAR}: Biogas recovery in the agriculture sector, {\natexlab{a}}.
\newblock URL \url{https://www.epa.gov/agstar}.
\newblock Library Catalog: www.epa.gov.

\bibitem[Laboratory()]{argonne_national_laboratory_renewable_nodate}
Argonne~National Laboratory.
\newblock Renewable natural gas database.
\newblock URL
  \url{https://www.anl.gov/es/reference/renewable-natural-gas-database}.
\newblock Library Catalog: www.anl.gov.

\bibitem[{CalRecycle}()]{calrecycle_california_2019}
{CalRecycle}.
\newblock California anaerobic digestion projects accepting organics from the
  municipal solid waste stream.

\bibitem[{US}~{EPA}({\natexlab{b}})]{us_epa_landfill_2016}
{OAR} {US}~{EPA}.
\newblock Landfill methane outreach program ({LMOP}), {\natexlab{b}}.
\newblock URL \url{https://www.epa.gov/lmop}.
\newblock Library Catalog: www.epa.gov.

\bibitem[Ong et~al.()Ong, Williams, and Kaffka]{ong_comparative_2014}
Matthew~D Ong, Robert~B Williams, and Stephen~R Kaffka.
\newblock Comparative assessment of technology options for biogas clean-up.
\newblock page 161.

\bibitem[Sun et~al.()Sun, Li, Yan, Liu, Yu, and Yu]{sun_selection_2015}
Qie Sun, Hailong Li, Jinying Yan, Longcheng Liu, Zhixin Yu, and Xinhai Yu.
\newblock Selection of appropriate biogas upgrading technology-a review of
  biogas cleaning, upgrading and utilisation.
\newblock 51:\penalty0 521--532.
\newblock ISSN 1364-0321.
\newblock \doi{10.1016/j.rser.2015.06.029}.
\newblock URL
  \url{http://www.sciencedirect.com/science/article/pii/S1364032115006012}.

\bibitem[Esposito et~al.()Esposito, Dellamuzia, Moretti, Fuoco, Giorno, and
  Jansen]{esposito_simultaneous_2019}
Elisa Esposito, Loredana Dellamuzia, Ugo Moretti, Alessio Fuoco, Lidietta
  Giorno, and Johannes~C. Jansen.
\newblock Simultaneous production of biomethane and food grade {CO}2 from
  biogas: an industrial case study.
\newblock 12\penalty0 (1):\penalty0 281--289.
\newblock ISSN 1754-5706.
\newblock \doi{10.1039/C8EE02897D}.
\newblock URL
  \url{https://pubs.rsc.org/en/content/articlelanding/2019/ee/c8ee02897d}.
\newblock Publisher: The Royal Society of Chemistry.

\bibitem[noa({\natexlab{a}})]{noauthor_natcarbatlas_nodate}
{NATCARB}/{ATLAS}, {\natexlab{a}}.
\newblock URL
  \url{https://www.netl.doe.gov/coal/carbon-storage/strategic-program-support/natcarb-atlas}.
\newblock Library Catalog: www.netl.doe.gov.

\bibitem[noa({\natexlab{b}})]{noauthor_project_nodate}
Project {OSRM}, {\natexlab{b}}.
\newblock URL \url{http://project-osrm.org/}.

\bibitem[Parker et~al.()Parker, Williams, Dominguez-Faus, and
  Scheitrum]{parker_renewable_2017}
Nathan Parker, Robert Williams, Rosa Dominguez-Faus, and Daniel Scheitrum.
\newblock Renewable natural gas in california: An assessment of the technical
  and economic potential.
\newblock 111:\penalty0 235--245.
\newblock ISSN 0301-4215.
\newblock \doi{10.1016/j.enpol.2017.09.034}.
\newblock URL
  \url{http://www.sciencedirect.com/science/article/pii/S0301421517305955}.

\bibitem[Tittmann et~al.()Tittmann, Parker, Hart, and
  Jenkins]{tittmann_spatially_2010}
P.~W. Tittmann, N.~C. Parker, Q.~J. Hart, and B.~M. Jenkins.
\newblock A spatially explicit techno-economic model of bioenergy and biofuels
  production in california.
\newblock 18\penalty0 (6):\penalty0 715--728.
\newblock ISSN 0966-6923.
\newblock \doi{10.1016/j.jtrangeo.2010.06.005}.
\newblock URL
  \url{http://www.sciencedirect.com/science/article/pii/S0966692310000864}.

\bibitem[Psarras et~al.()Psarras, Comello, Bains, Charoensawadpong,
  Reichelstein, and Wilcox]{psarras_carbon_2017}
Peter~C. Psarras, Stephen Comello, Praveen Bains, Panunya Charoensawadpong,
  Stefan Reichelstein, and Jennifer Wilcox.
\newblock Carbon capture and utilization in the industrial sector.
\newblock 51\penalty0 (19):\penalty0 11440--11449.
\newblock ISSN 1520-5851.
\newblock \doi{10.1021/acs.est.7b01723}.

\bibitem[{McCollum} and Ogden()]{mccollum_techno-economic_2006}
David~L. {McCollum} and Joan~M. Ogden.
\newblock Techno-economic models for carbon dioxide compression, transport, and
  storage \&amp; correlations for estimating carbon dioxide density and
  viscosity.
\newblock URL \url{https://escholarship.org/uc/item/1zg00532}.

\bibitem[Trautz et~al.()Trautz, Swisher, Chiaramonte, Hollis, Perron, Pronske,
  Myhre, Stone, Saini, Jordan, Wagoner, and Kent]{trautz_california_2018}
Robert Trautz, Joseph Swisher, Laura Chiaramonte, Rebecca Hollis, Joshua
  Perron, Keith Pronske, Richard Myhre, Marian Stone, Dayanand Saini, Preston
  Jordan, Jeff Wagoner, and Ronald Kent.
\newblock California {CO}2 storage assurance facility enterprise (c2safe):
  Final technical report.
\newblock URL
  \url{https://www.osti.gov/biblio/1452864-california-co2-storage-assurance-facility-enterprise-c2safe-final-technical-report}.

\bibitem[Li et~al.()Li, Zhang, Liu, Chen, He, and Liu]{li_comparison_2013}
Yeqing Li, Ruihong Zhang, Guangqing Liu, Chang Chen, Yanfeng He, and Xiaoying
  Liu.
\newblock Comparison of methane production potential, biodegradability, and
  kinetics of different organic substrates.
\newblock 149:\penalty0 565--569.
\newblock ISSN 0960-8524.
\newblock \doi{10.1016/j.biortech.2013.09.063}.
\newblock URL
  \url{http://www.sciencedirect.com/science/article/pii/S0960852413014958}.

\bibitem[{US}~{EPA}({\natexlab{c}})]{us_epa_emissions_2015}
{OAR} {US}~{EPA}.
\newblock Emissions \& generation resource integrated database ({eGRID}),
  {\natexlab{c}}.
\newblock URL
  \url{https://www.epa.gov/energy/emissions-generation-resource-integrated-database-egrid}.
\newblock Library Catalog: www.epa.gov.

\bibitem[Fund()]{environmental_defense_fund_green_nodate}
Environmental~Defense Fund.
\newblock Green freight math: How to calculate emissions for a truck move.
\newblock URL
  \url{https://business.edf.org/insights/green-freight-math-how-to-calculate-emissions-for-a-truck-move/}.
\newblock Library Catalog: business.edf.org.

\bibitem[EIA(2021)]{eia2020rng}
EIA.
\newblock California natural gas consumption by end use.
\newblock \url{https://www.eia.gov/dnav/ng/ng_cons_sum_dcu_SCA_a.htm}, 2021.

\bibitem[N-79-20(2020)]{execorder}
California Executive~Order N-79-20, 2020.

\bibitem[Burns and Jacobson()]{burns_jacobson_2021}
Erin Burns and Rory Jacobson.
\newblock Enhancing and expanding the 45q tax credit for direct air capture.

\bibitem[McQueen et~al.(2020)McQueen, Psarras, Pilorg{\'e}, Liguori, He, Yuan,
  Woodall, Kian, Pierpoint, Jurewicz, et~al.]{mcqueen2020cost}
Noah McQueen, Peter Psarras, H{\'e}l{\`e}ne Pilorg{\'e}, Simona Liguori, Jiajun
  He, Mengyao Yuan, Caleb~M Woodall, Kourosh Kian, Lara Pierpoint, Jacob
  Jurewicz, et~al.
\newblock Cost analysis of direct air capture and sequestration coupled to
  low-carbon thermal energy in the united states.
\newblock \emph{Environmental science \& technology}, 54\penalty0
  (12):\penalty0 7542--7551, 2020.

\bibitem[Sanchez et~al.({\natexlab{b}})Sanchez, Zimring, Mater, Harrell,
  Kelley, Muszynski, Edwards, Smith, Monper, Marley, and
  Russer]{sanchez_literature_2020}
Daniel~L. Sanchez, Teal Zimring, Catherine Mater, Katie Harrell, Stephen
  Kelley, Lech Muszynski, Ben Edwards, Samantha Smith, Kyle Monper,
  {AnnaClaire} Marley, and Max Russer.
\newblock Literature review and evaluation of research gaps to support wood
  products innovation.
\newblock {\natexlab{b}}.

\bibitem[Mahone et~al.()Mahone, Kahn-Lang, Li, Ryan, Subin, Allen, De~Moor, and
  Price]{mahone_deep_nodate}
Amber Mahone, Jenya Kahn-Lang, Vivian Li, Nancy Ryan, Zachary Subin, Douglas
  Allen, Gerrit De~Moor, and Snuller Price.
\newblock Deep decarbonizaiton in a high renewables future.
\newblock URL
  \url{https://www.ethree.com/wp-content/uploads/2018/06/Deep_Decarbonization_in_a_High_Renewables_Future_CEC-500-2018-012-1.pdf}.

\bibitem[Institue()]{gas_technology_institue_low_nodate}
Gas~Technology Institue.
\newblock Low carbon renewable natural gas ({RNG}) from wood wastes.
\newblock URL
  \url{https://www.gti.energy/wp-content/uploads/2019/02/Low-Carbon-Renewable-Natural-Gas-RNG-from-Wood-Wastes-Final-Report-Feb2019.pdf}.

\bibitem[Grubert()]{grubert_at_2020}
Emily Grubert.
\newblock At scale, renewable natural gas systems could be climate intensive:
  The influence of methane feedstock and leakage rates.
\newblock ISSN 1748-9326.
\newblock \doi{10.1088/1748-9326/ab9335}.
\newblock URL \url{http://iopscience.iop.org/10.1088/1748-9326/ab9335}.

\end{thebibliography}
\end{singlespace}
\end{document}